\documentclass[a4,11pt,epsf,amssymb]{article}
\usepackage{amsmath,amsfonts,latexsym,amssymb, color}
\begin{document}

\title{ \bf Quantile Estimation of A General Single-Index Model }
\author{ Efang Kong\thanks{%
Eurandom, Technische Universiteit Eindhoven, The Netherlands. %
  E-mail address: \texttt{kong@eurandom.tue.nl.}}\\
%EndAName
{\normalsize \textit{Technische Universiteit Eindhoven}}\\ \\
Yingcun Xia\thanks{%
Department of Statistics and Applied Probability, National
University of Singapore, Singapore.
http://www.stat.nus.edu.sg/\symbol{126}staxyc.
 E-mail address: \texttt{staxyc@nus.edu.sg.}}\\
{\normalsize \textit{National University of Singapore, Singapore}} }
\date{}
\def\beginn{\begin{eqnarray*}}
\def\endn{\end{eqnarray*}}
\def\beginy{\begin{eqnarray}}
\def\endy{\end{eqnarray}}
\def\begine{\begin{equation}}
\def\ende{\end{equation}}
\def\n{\nonumber}
\newtheorem{Theorem}{Theorem}[section]
\newtheorem{Example}[Theorem]{Example}
\newtheorem{Lemma}[Theorem]{Lemma}
\newtheorem{Note}[Theorem]{Note}
\newtheorem{Proposition}[Theorem]{Proposition}
\newtheorem{Corollary}[Theorem]{Corollary}
\newtheorem{Remark}[Theorem]{Remark}
\newcommand{\un}[1]{\underline{#1}}

\def\X{{\mathbf X}}
\def\x{\textit{x}}
\def\Cov{\mbox{\rm Cov}}
\def\Var{\mbox{\rm Var}}
\def\YX{_{Y|X}}
\def\argmin{\mbox{\rm arg}\min}
\def\b{{\mathbf b}}
\def\e{{\varepsilon}}
\def\F{{\mathbf F}}
\def\be{\mbox{ e}}
\def\bu{\mbox{ U}}
\def\bv{\mbox{ V}}
\def\CV{\mbox{\small CV}}
\def\cov{\mbox{ cov}}
\def\T{\mbox{\rm T}}
\def\L{\mbox{\rm L}}
\def\argmax{\mbox{\rm argmax}}

\def\pperp{\perp\hspace{-.25cm}\perp}
\def\t{{\hspace{-0.05cm}\top}}
\def\A{{\cal A}}
\def\Bt{\theta^\top}
\def\E{{\cal E}}
\def\D{{\cal D}}
\def\N{{\cal N}}
\def\M{{{\cal M}}}
\def\R{{\cal R}}
\def\rt{\raisebox{1.5ex}[0pt]}
\def\btd{\bigtriangledown}
\def\btdt{\bigtriangledown^\t\hspace{-0.1cm}}
\def\dfor{{\qquad \mbox{\rm for} \quad}}
\def\C{{\cal C}}
\def\g{{\sl g}}
\def\rt{\raisebox{1.8ex}[0pt]}
\def\rtf{\raisebox{1.2ex}[0pt]}
\def\ditem{\vspace{-0.2cm} \item}

\maketitle

\begin{quotation}

\baselineskip1.8em

{\bf Abstract} The single-index model is one of the most popular
semiparametric models in Econometrics. In this paper, we define a
quantile regression single-index model, which includes the
 single-index structure for conditional mean and for  conditional
 variance.


\end{quotation}

\baselineskip1.8em

\def\x{{\bf x}}

\section{Introduction}

Regression quantiles, along with the dual methods of regression rank
scores, can be considered one of the major statistical breakthroughs
of the past decades. Its advantages over the other estimation
methods have been well investigated. Regression quantile methods
provide a much more complete statistical analysis of the stochastic
relationships among variables; in addition, they are more robust
against possible outliers or extreme values, and can be computed via
traditional linear programming methods. Although median regression
ideas go back to the 18th century and the work of Laplace,
regression quantile methods were first introduced by Koenker and
Bassett (1978). The linear regression quantile is very useful, but
like linear regression it is not flexible enough to capture
complicated relations. For quantile regression, this disadvantage is
even worse. As an example, consider the popular AR(1)-ARCH(1) model:
\beginn
&&y_t = \alpha_0 + \alpha_1 y_{t-1} + \e_t,\ \e_t=\sigma_tz_t,\ z_t\sim \mbox{ IID}\\
&& \sigma_t^2=\beta_0+\beta_1\e^2_{t-1},\ \beta_0>0,\ \beta_1\ge 0,
\endn
which cannot be fitted well by the linear quantile model.

In this paper, we focus on an important special case when the loss
function is specified as
\beginy
  \rho_\tau(v) = \tau I(v>0) v + (\tau-1) I(v\le 0) v,\label{bath}
\endy
where $ 0<  \tau < 1 $ and $ I(.) $ is the identity function,
leading to the $\tau$th quantile regression, see Koenker and Bassett
(1978).

In a nonparametric setting, we can state the problem as follows.
Suppose $Y$ is the response variable and $X\in R^d$ are the
covariates. For loss function $\rho_\tau(.)$, we are interested in a
function  $ m_\tau(x) $, such that
\beginy
m_\tau(x) = \mbox{arg}\min E\{\rho_\tau[Y-m(X)]\Big|X=x\}\quad
\mbox{with respect to }  m(.)\in L_1.\label{yy}
\endy
The function $ m_\tau(x) $ is called the $\tau-$th quantile
nonparametric regression function of $ Y $ on $ X$. The application
of nonparametric quantile estimation has been intensively
investigated in the literature. See for example  Koenker (2005)
 and Kong et al (2008). As in nonparametric
estimation of the conditional mean function, there is the ``curse of
dimensionality'' in estimating the typically multivariable function
$ m_\tau(.) $. The dimension reduction approach can thus be applied
here, by considering
\beginy
m_\tau(\theta^\top x) = \mbox{arg}\min E\{\rho_\tau(Y-m(\theta^\t
{X}))|{X}={x}\}\quad \mbox{with respect to }\ \theta \in \Theta\
\mbox{and}\  m(.)\in L_1,\label{yy}
\endy
where $\Theta=\{\theta:|\theta|=1\}$. Ideally, we come to a
single-index quantile model
\beginy
Y=m(\theta_0^\t {X})+\e,\quad E(\varphi(\e)|{X})=0,\ a.s.\label{sim}
\endy
where $\varphi(.)$ is the piecewise derivative function of $\rho(.)$ in (\ref{bath}).
A typical model is the general single-index model,
$$
 Y = g(\theta_0^\top X, \varepsilon)
$$
where $ \varepsilon $ is independent of $ X $. Under such a model
specification, it is easy to see that
$$
m_\tau(x) = g_\tau(\theta^\top x) \equiv \min_{v}\{v:
P(g(\theta_0^\top x, \varepsilon) \le v) \ge \tau \}.
$$
For the conditional heteroscadiscity model, where $ g(\theta_0^\top
X, \varepsilon) = g(\theta_0^\top X) \varepsilon $, we even have
$$
 m_\tau(x) = g(\theta_0^\top X) Q_\tau(\varepsilon)
$$
where $ Q_\tau(\varepsilon)$ is the $\tau-$th quantile of $
\varepsilon$. An interesting special case for this setting is the
ARCH(p) model, where  $ X = (y_{t-1}^2, ..., y_{t-p}^2)^\top $ and $
 Y = y_t $ in a time series setting.

%\begin{Proposition} \rm For the general single-index model, we have
%$$
%g_q(\theta^\top x) = \mbox{arg}\min_{ Q\in \R}
%$$
%\end{Proposition}

Our main focus is the estimation of $\theta_0$. Suppose
$\{{X}_i,Y_i\}_{i=1}^n$ are I.I.D. observations from underlying
model (\ref{sim}). We propose to estimate the index parameter
$\theta_0$ by
\beginy
\hat\theta=\argmin\limits_{\theta\in \Theta}\ \min\limits_{a_j,b_j}\
\sum\limits_{i=1}^n\sum\limits_{j=1}^n K(\theta^\t
{X}_{ij}/h)\rho(Y_i-a_j-b_j\theta^\t {X}_{ij}),\quad
{X}_{ij}={X}_{i}-{X}_{j}\label{Perry}
\endy
where  $K(.)$ is a kernel function and $h$ is a bandwidth. The
minimization in (\ref{Perry}) can be realized through iteration.
First for any initial estimate $\vartheta\in \Theta$, denote by
$[\hat a_{\vartheta}({x}),\hat b_{\vartheta}({x})]$, the minimizer
of
\beginy
 \sum\limits_{i=1}^n
K(\vartheta^\top {X}_{ix}/h)\rho(Y_i-a-b\vartheta^\t
{X}_{ix})\quad\mbox{with respect to } a \mbox{ and } b, \label{ja}
\endy
 where ${X}_{ix}={X}_i-{x}$. The estimate of $\theta_0$ is then updated
 by
\beginy
\hat\theta=\argmin\limits_{\theta\in \Theta}\
 \sum\limits_{i=1}^n\sum\limits_{j=1}^n
K(\vartheta^\t {X}_{ij}/h)\rho\{Y_i-\hat a_{\vartheta}({X}_j)-\hat
b_{\vartheta}({X}_j)\theta^\t {X}_{ij}\}.\label{hei}
\endy
 Repeat (\ref{ja}) and (\ref{hei}) until convergence. The true value
$\theta_0$ is thus estimated by the standardized  final estimate
 $\hat\theta:=\hat\theta/|\hat\theta|$.

\section{Numerical studies}

Again, the calculation of the above minimization problem can be
decomposed into two minimization problems.

\begin{itemize}

\item Fixing $\theta = \vartheta$ and $ w_{ij}^\vartheta =
K_h(\vartheta^\top X_{ij})$, the estimation of $ a_j $ and $ d_j $
are
\begin{eqnarray*}
  \sum_{i=1}^n \rho\{ Y_i - a_j - d_j \vartheta^\top X_{ij} \}
  w_{ij}^\vartheta.
\end{eqnarray*}

\item Fixing $ a_j $ and $ d_j $, the minimization with respect to $
\theta $ can be done as follows. Again, let
$$
 Y_{ij}^\vartheta = Y_i (w_{ij}^\vartheta)^{1/2} - a_j (w_{ij}^\vartheta)^{1/2} , \quad
 X_{ij}^\vartheta = d_j X_{ij} (w_{ij}^\vartheta)^{1/2}.
$$
Then the problem becomes
$$
\min_\vartheta \sum_{i,j=1}^n \rho\{ Y_{ij}^\vartheta - \theta^\top
 X_{ij}^\vartheta\}
$$
Suppose the solution to the above problem is $\theta $. Standardize
it to $ \theta := \theta/||\theta|| $.

\end{itemize}

Set $ \vartheta = \theta $ and repeat the two steps until
convergence. Note that both steps are simple linear quantile
regression problems and that several  efficient algorithms are
available, see Koenker (2005).

\begin{Example}[Single-index median regression] \rm Consider the following model
\beginy
    y = \exp\{-5(\theta_0^\top X)^2\} + \varepsilon, \label{qrex}
\endy
where $ X \sim \Sigma_0^{1/2} X_0 $ with $ X_0 \sim N(0, I_5) $ and
$ \Sigma_0 = (0.5^{|i-j|})_{0 \le i, j \le 5} $. For the noise term,
we consider several distributions with both heavy tail and thin
tails as well. For simplicity, we consider the median regression
only. As a comparison, we also run the MAVE where a least square
type estimation is used. With different sample sizes $ n = 100, 200
$, we carried out 100 replications. The calculation results are
listed in Table \ref{table1}.

\begin{table}[h]
\centering \caption{Estimation errors (and standard errors) for
model (\ref{qrex}) based on quadratic loss function and 50\%
quantiles  }\label{table1}
\begin{tabular}{|c|c|c|c|c|c|}
\hline
 &  & \multicolumn{4}{c|}{Distribution of $
\varepsilon$}
\\
\cline{3-6}
 size & method & $0.05 t(1)$ & $0.1(N(0,1)^4-3)$ & $\sqrt{5}t(5)/20$ & N(0,1)/4
\\
\hline
 100 & MAVE   &  0.3641(0.3526)   &  0.3530(0.3102) &   0.0401(0.0182)& 0.0581(0.0263) \\
     & qMAVE  &  0.0902(0.1074)   &  0.1512(0.1957) &  0.0833(0.0785) & 0.1146(0.0651)\\
\hline
 200 & MAVE   & 0.3381(0.3389)    &  0.2859(0.2887) &   0.0232(0.0091) & 0.0373(0.0147)\\
     & qMAVE  &  0.0681(0.1415)   &  0.0581(0.0698) &  0.0402(0.0173) & 0.0652(0.0272)\\
\hline
\end{tabular}
\end{table}

The MAVE method with quadratic loss function has very bad
performance when the noise has heavy tail (e.g. $t(1)$) or is highly
asymmetric (e.g. $ N(0,1)^4$). With the absolute value loss
function, the performance is much better. Even in the situation when
the noise has thin tail and symmetric, qMAVE still performance
reasonably well.

\end{Example}

\section{Assumptions and asymptotic properties }
We adopt model (\ref{sim}) throughout and make the additional
assumption that  $\{(X_i,Y_i)\}_{i=1}^\infty$ are I.I.D.
observations. The extension to the case of weakly dependent time
series  should be straightforward but complicates matters without
adding anything conceptually. Furthermore, the following conditions
are assumed in the proofs of Theorem \ref{T0}.
\begin{description}

\item{(A1)} For each $v\in \R,\ \rho(v)$ is absolutely continuous,
i.e., there is a function $\varphi(.)$ such that $
\rho(v)=\rho(0)+\int_0^v\varphi(t)dt. $ The probability density
function of $\e_i$ is bounded and  continuously differentiable. $
E\{\varphi(\e_i)| {X}_i\}=0 $ almost surely and
$E|\varphi(\e_i)|^{\nu_1}\le M_0<\infty$ for some $\nu_1>2.$

\item{(A2)} Function $\varphi(.)$ satisfies the
Lipschitz condition  in $(a_j,a_{j+1}),\ j=0,\cdots,m$, where $a_1
<\cdots < a_m$ are finite number of jump discontinuity points of
$\varphi(.)$,  $a_0\equiv-\infty$,  $a_{m+1}\equiv+\infty$ and $ m <
\infty$.

\item{(A3)}  Kernel function $K(.)$ is symmetric density function with a compact support and satisfies
$|{u}^{ {j}}K( {u})-{v}^{ {j}}K( {v})|\le C|u-v| $
 for all $j$ with $0\le j \le 3$.

\item{(A4)} The link function $m(.)$ defined in (\ref{sim}) has
continuous and bounded derivatives up to the third order.

\item{(A5)} The smoothing parameter $h$ is chosen such that $nh^4\to \infty$ and
$nh^5/\log n<\infty$.

%\item{(A5)} The uniform mixing coefficient satisfies that $\phi(m)=m^{-k}$ for some $k>0$,
% and $n^{k/10}=O\{(nh)^{1/2}\tau_n\}$, where
%$\tau_n=(nh/\log, n)^{-3/4}$.
\end{description}
Note that (A1) and (A2)  are satisfied in quantile regression with $
\rho(.) = \rho_\tau(.) $ given in (\ref{bath}). Condition (A3) and
(A4) are standard in kernel smoothing. Based on (A1) and (A2), Hong
(2003) proved that
 there is a constant $C>0$, such that for all small
$t$ and all $x$,
\beginy
E\Big[\{\varphi(Y-t-a)-\varphi(Y-a)\}^2|X=x\Big]\le
C|t|\label{A81}
\endy
holds for all  $(a,x)$ in a neighborhood of $\{m(x^\t\theta_0),x\}$.
Define
\beginy
G(t;{x})=E\{\rho\{Y-m(x^\t\theta_0)+t\}|{X}={x}\},\quad
G_i(t,{x})=(\partial^i/\partial t^i)G(t;{x}),\ i=1,2,3.\label{kk}
\endy
Then it follows that
$$g(x)\stackrel{def}{=}G_2(0;x)\ge C>0$$
and $ G_3(t,{x}) $ is continuous and uniformly bounded for all $
x\in {\mathcal D}\mbox{ and }t \mbox{ near }0 $.  For quantile
regression, $g(x)=f_\e(0|x)$, where $f_\e(.|x)$ is the conditional
probability density function of $\e$  given $X=x$.

\section{Initial estimator of $\theta_0$}

We use the average derivative estimation (ADE, H\"ardle and Stocker,
1989; Chaudhuri et al., 1997) method to obtain an initial estimate
of $\theta_0$, by observing the fact that $ E[\partial m(\theta_0^\t
X)/\partial X]=\theta_0 E[\partial m(\theta_0^\t X)/\partial
(\theta_0^\t X)] $ and
\begin{equation}
\theta_0 =  E[\partial m(\theta_0^\t X)/\partial X]/ E[\partial
m(\theta_0^\t X)/\partial (\theta_0^\t X)].  \label{initial0}
\end{equation}
For any ${x}\in R^d$ and a kernel density function $H(.): R^d\to R^+$, denote by $[\hat a({x}),\hat b({x})]$,  the
minimizer of the following quantity
 \beginn
 \sum\limits_{i=1}^n
H( {X}_{ix}/h_0)\rho(Y_i-a-b^\t {X}_{ix}),
\endn
with respect to $ a$ and  $b$.  Observing (\ref{initial0}), an initial
estimate of $\theta_0$ could be constructed as follows
\beginy
\vartheta=\sum_{j=1}^n {\color{red} c(X_j)} \hat
b({X}_j)\Big/\Big|\sum_{j=1}^n  {\color{red} c(X_j)} \hat
b({X}_j)\Big|, \label{lisa}
\endy
 where $ C(x) $ is some trimming function introduced to deal with boundary effects.

The consistency of
$\vartheta$ in  (\ref{lisa}) can be proved using the results
on the uniform Bahadur representation of
$\hat b({x})$ over any compact subset $\mathcal D$ of the support of $X$.
Suppose $H(.)$ is symmetric about $0$ in each coordinate direction and the conditions
 in Proposition 3.1 and Corollary 3.3 in Kong et al
(2007) are met, especially $nh_0^{d+4}/\log n<\infty$ and $nh_0^{d}/\log n\to \infty$.
Then with probability one,
\beginy
\hat b({x})=m'(\theta_0^\t
{x})\theta_0+\frac{1}{nh_0^{d+1}\{fg\}(x)}\sum\limits_{i=1}^{n}
H({X}_{ix}/h_0)\varphi(\e_i)
{X}_{ix}/h_0
+O\Big\{h_0^{-1}\Big(\frac{\log n}{nh_0^d}\Big)^{3/4}\Big\}
\endy
uniformly in $x\in \mathcal D,$
where $\{fg\}(x)=f(x)g(x)$ with $f(.)$  the density function of $X$ and
$g(x)>0$  some deterministic function.
  This in turn implies that with
probability one,
\beginn
\frac{1}{n}\sum_{j=1}^n {\color{red} c(X_j)} \hat
b({X}_j)&=&m'(\theta_0^\t
{x})\theta_0+\frac{1}{n^2h_0^{d+1}}\sum\limits_{i,j=1}^{n}
{\color{red} c(X_j)} \{fg\}^{-1}({X}_j)
H({X}_{ij}/h_0)\varphi(\e_i)
{X}_{ij}/h_0\\
&& +O\Big\{h_0^{-1}\Big(\frac{\log n}{nh_0^d}\Big)^{3/4}\Big\}.
\endn
Using results in Masry (1996), we know that with
probability 1,
\beginn
\frac{1}{nh_0^d}\sum\limits_{i=1}^{n} H({X}_{ix}/h_0)\varphi(\e_i)
\frac{{X}_{ix}}{h_0}=O\{(nh_0^d/\log n)^{-1/2}\}
\endn
uniformly in $x\in \mathcal D,$ whence
\beginn
\frac{1}{n^2h_0^{d+1}2}\sum\limits_{i,j=1}^{n} {\color{red} c(X_j)}\{fg\} ^{-1}({X}_j)
H({X}_{ij}/h_0)\varphi(\e_i)
\frac{{X}_{ij}}{h_0}=O\{h_0^{-1}(nh_0^d/\log n)^{-1/2}\}
\endn
almost surely. Therefore, concerning the initial
estimator $\vartheta$ in (\ref{lisa}), we have
\beginy
\delta_\vartheta\equiv\theta_0-\vartheta=O\{h_0^{-1}(nh_0^d/\log
n)^{-1/2}\}\label{hong}
\endy
almost surely. Consequently from now on, we focus on parametric
space $\Theta_n\equiv \{\vartheta: |\delta_\vartheta|<
Ch(nh_0^{d+2}/\log n)^{-1/2}\}$ for some constant $C>0$.

\section{Asymptotics of $\hat a_{\vartheta}({x})$ and $\hat b_{\vartheta}({x})$}

 For any $\vartheta\in \Theta_n$, denote by $f_{\vartheta}({x}) $ and $ F_{\vartheta}({x})$
,the probability density function and  distribution function of
$\vartheta^\t {X}$ at $\vartheta^\t {x}$ respectively, and for any $v\in R$ and $x\in \mathcal D\subset R^d$, define
\beginn
&&m_\vartheta(v)=\arg\min_a E\{\rho(Y-a)|X^\t\vartheta=v\},\\
&&G_\vartheta(t,x)=E\{\rho(Y-m_\vartheta(\vartheta^\t
x)+t)|\vartheta^\t X=\vartheta^\t
x\},\\
&& G^i_\vartheta(t,x)=(\partial^i/\partial t^i)G_\vartheta(t,x),\
i=1,2;\quad g_\vartheta(x)=G^2_\vartheta(m_\vartheta(x),x)
\endn
Apparently $g_{\theta_0}(x)\equiv g(x)$.
We assume that for any $\vartheta$ in a neighborhood of $\theta_0,$
$G^2_\vartheta(t,x)$ is continuous and uniformly bounded in
the neighborhood of $(m_\vartheta(x),x)$ and there exists some $\delta>0$ such that
$g_\vartheta(x)>\delta$ for $\vartheta$ near enough $\theta_0$ and $x\in \mathcal D.$

\noindent With initial estimate $\vartheta,$ let $[\hat
a_j,\hat b_j]\equiv [\hat a_\vartheta(X_j),\hat b_\vartheta(X_j)]$
be  the solution to (\ref{ja}) with $x$ specified as $X_j$.
If the smoothing parameter $h$ is chosen such that $nh/\log n\to \infty$ and $nh^5/\log n<\infty,$
using the results on  uniform
Bahadur representation  in Kong et al (2007), we have
\beginy
&&\hspace{-.5cm}\hat
a_j-m_\vartheta(X_j)=\frac{1}{nh}\{g.f\}^{-1}_\vartheta(X_j)
\sum\limits_{i=1}^{n} K^\vartheta_{ij}\varphi(Y^*_{ij})
+O\Big\{\Big(\frac{\log n}{nh}\Big)^{3/4}\Big\},\label{cut}\\
\n&&\hspace{-.5cm} h\{\hat
b_j-m'_\vartheta(X_j)\}=\frac{1}{nh}\{g.f\}^{-1}_\vartheta(X_j)\sum\limits_{i=1}^{n}
K^\vartheta_{ij}\varphi(Y^*_{ij})
X_{ij}^\t\vartheta/h+O\Big\{\Big(\frac{\log n}{nh}\Big)^{3/4}\Big\},
\endy
uniformly in $X_j\in \mathcal D$, where $
K^\vartheta_{ij}=K(X_{ij}^\t\vartheta/h)$,
$Y_{ij}^*=Y_{i}-m_\vartheta(X_j)-m'_\vartheta(X_j)X_{ij}^\t\vartheta$
 and $\{g.f\}_\vartheta(.) = g_\vartheta(.) f_\vartheta(.) $.
Note that  $m_\vartheta(X_j)\stackrel{def}{=}m_\vartheta(X_j^\t\vartheta)$ and
 $m'_\vartheta(X_j)\stackrel{def}{=}m'_\vartheta(X_j^\t\vartheta)$.

Combined with Lemma \ref{LL0} and Lemma \ref{great} in the Appendix, further to (\ref{cut}), we have
\beginy
\hspace{-.5cm}\n\hat
a_j-a_j&=&\frac{1}{2}m^{''}(X_j^\t\theta_0)
_\vartheta(X_j)h^2+b_j\delta_\vartheta^\t\{(\nu/\mu)_\vartheta(X_j)-X_j\}\\
&& + (nh)^{-1}\{gf\}^{-1}_\vartheta(X_j) \sum\limits_{i=1}^{n}
\varphi_{ij} +O\Big\{\Big(\frac{\log
n}{nh}\Big)^{3/4}+h^4+h\delta_\vartheta\Big\},\label{lu1}\\
\n \hat b_j-b_j&=&
h^2\Big[\frac{1}{2}m^{''}(X_j^\t\theta_0)\{(f\mu)'/(fg)\}_\vartheta(X_j)
+\frac{1}{6}m^{(3)}(X_j^\t\theta_0)\{(f\mu)/(fg)\}_\vartheta(X_j)\Big]\\
\n&&
+b_j\delta_\vartheta^\t\{(\mu\nu'-\mu'\nu)/\mu^2\}_\vartheta(X_j)+(nh^2)^{-1}\{gf\}^{-1}_\vartheta(X_j)
\sum\limits_{i=1}^{n} \tilde \varphi_{ij}\\
\n && +O\Big\{h^4+h^2\delta_\vartheta+\Big(\frac{\log
n}{nh}\Big)^{3/4}/h\Big\}
\endy
uniformly in $j$ with $ X_j \in {{\cal D}}$, where $(\nu/\mu)_\vartheta(X_j)\equiv
\nu_\vartheta(X_j^\t\vartheta)/\mu_\vartheta(X_j^\t\vartheta)$,
\beginy
\mu_\vartheta(v)=E[g(X)|X^\t\vartheta=v],\quad
\nu_\vartheta(v)=E[g(X)X|X^\t\vartheta=v].\label{yyy}
\endy
 and $\varphi_{ij}$ and $ \tilde \varphi_{ij}$ are zero-mean
I.I.D. random variables defined as
\beginy
&&\hspace{-.5cm}\varphi_{ij}=K^\vartheta_{ij}\varphi(Y^*_{ij})
-E[K^\vartheta_{ij}\varphi(Y^*_{ij})],\label{lu2}\\
\n&&\hspace{-.5cm}\tilde \varphi_{ij}=K^\vartheta_{ij}\varphi(Y^*_{ij})X_{ij}^\t\vartheta/h
-E[K^\vartheta_{ij}\varphi(Y^*_{ij})X_{ij}^\t\vartheta/h].
\endy
Note that (\ref{lu1}) focuses on the almost sure property of $[\hat a_j,
\hat b_j].$
 Welsh (1996) studied their the asymptotic bias and variance,
i.e.
\beginy
\n &&E\{\hat a(x)\}=m_\vartheta(\vartheta^\t x)+O(h^2),\quad E\{\hat b(x)\}=m'_\vartheta(\vartheta^\t x)+O(h^2),\\
 &&\Var\{ \hat
a(x)\}=O(n^{-1}h^{-3}),\quad \Var\{ \hat
b(x)\}=O(n^{-1}h^{-3}),\label{welsh}
\endy
and the $O(.)$s are uniformly in $x$ in any compact subset of the support of $X$.

\section{Asymptotics of $\hat \theta$}

For the previously obtained $\vartheta$, $\hat a_j,\ \hat b_j,\
j=1,\cdots,n$, suppose $\hat\theta$ minimizes
$\tilde\Phi_{n}(\theta)$, where
\beginn
\sum\limits_{i=1}^n\sum\limits_{j=1}^n
K^\vartheta_{ij}\rho(Y_i-\hat a_j-\hat b_j\theta^\t {X}_{ij}) +\frac{n^2h}{2}(\theta-\vartheta)^\t
\vartheta\vartheta^\t (\theta-\vartheta).
\endn
Apparently, $\hat\theta$ also minimizes
\beginy
\n&&\tilde\Phi_n(\theta)=\Phi_n(\theta)+{n^2h}\{\frac{1}{2}(\theta-\theta_0)^\t
\vartheta\vartheta^\t (\theta-\theta_0)+(\theta_0-\vartheta)^\t
\vartheta\vartheta^\t(\theta-\theta_0)\}\\
&&\Phi_n(\theta)=\sum\limits_{i=1}^n\sum\limits_{j=1}^n
K^\vartheta_{ij}\{\rho(Y_i-\hat a_j-\hat b_j\theta^\t
{X}_{ij})-\rho(Y_{ij})\},\label{rev}
\endy
where $Y_{ij}\equiv
Y_i-\hat a_j-\hat b_j{X}_{ij}^\t\theta_0$. Let
$a_{n\vartheta}=\max\{(n\log\log n)^{-1/2},|\delta_\vartheta|\}$.  As $|\vartheta-\theta_0|=O( a_{n\vartheta})$,
$\vartheta\vartheta^\t=\theta_0\theta_0^\t+O(a_{n\vartheta})$, whence for
 any $\theta$ with $\delta_\theta\stackrel{def}{=}\theta_0-\theta=O( a_{n\vartheta})$, we have
\beginn
\tilde\Phi_n(\theta)=\Phi_n(\theta)+{n^2h}\{\frac{1}{2}\delta_\theta^\t
\theta_0\theta_0^\t \delta_\theta-\delta_\vartheta^\t
\theta_0\theta_0^\t\delta_\theta\}+o(n^2ha_{n\vartheta}^2).
\endn
 Write
$\Phi_n(\theta)=E[\Phi_n(\theta)]+\delta_\theta^\t\{R_{n1}(\theta)-ER_{n1}(\theta)\}+R_{n2}(\theta)-ER_{n2}(\theta),$
 where
\beginn
R_{n1} =\sum\limits_{i,j}K^\vartheta_{ij}\varphi(Y_{ij})\hat b_j
X_{ij},\
R_{n2}(\theta)=\sum\limits_{i,j}K^\vartheta_{ij}\Big[\rho(Y_i-\hat
a_j-\hat b_j\theta^\t {X}_{ij})-\rho(Y_{ij})-\delta_\theta^\t
\varphi(Y_{ij})\hat b_jX_{ij}\Big].
\endn
 Applying the results on $E(\Phi_n(\theta))$ in Lemma \ref{gg}, we
have
\beginy
 \Phi_n(\theta)=\delta_\theta^\t
R_{n1}+\frac{1}{2}\delta_\theta^\t
G_{n\vartheta}\delta_\theta\{1+o(1)\}+R_{n2}(\theta)-ER_{n2}(\theta),\label{rachel}
\endy
where
\beginn
&&G_{n\vartheta}=\sum\limits_{i,j}E[K^\vartheta_{ij}g(X_i) \hat b_j^2
X_{ij}X_{ij}^\t]=n^2hS_2\{1+O(\delta_\vartheta)\},\\
&& S_2=\int \{m'({X}^\t\theta_0)\}^2
\omega_{\theta_0}({X})f_{\theta_0}(X)dX,
\endn
and $
\omega_{\vartheta}({x})=E\{g_\vartheta(X)(X-x)(X-x)^\t|X^\t\vartheta=x^\t\vartheta\}.$
Consequently,
\beginn
\tilde \Phi_n(\theta)=
(R_{n1}-\theta_0\theta_0^\t)\delta_\theta+\frac{1}{2}\delta_\theta^\t
(G_{n\vartheta}+n^2h\theta_0\theta_0^\t)\delta_\theta\{1+o(1)\}+R_{n2}(\theta)-ER_{n2}(\theta).
\endn

Our main result is as follows
\begin{Theorem}\label{T0} Suppose  (A1)-(A4) hold.
With $\nu_\vartheta(.)$ and $\mu_\vartheta(.)$ as defined in
(\ref{yyy}), we have
\beginy
\n\hat\theta-\theta_0&=&(S_2+\theta_0\theta_0^\t)^{-1}\frac{1}{n}\sum\limits_{i}\varphi(\e_i)b_i\{\varpi
f\}_{\theta_0}(X_i)\\
&&\n-(S_2+\theta_0\theta_0^\t)^{-1}(\Omega_{n\vartheta}+\theta_0\theta_0^\t)\delta_\vartheta+\alpha_n |\vartheta-\theta_0|  + o(n^{-1/2})\\
\n&=&(S_2+\theta_0\theta_0^\t)^{-1}\frac{1}{n}\sum\limits_{i}\varphi(\e_i)b_i\{\varpi
f\}_{\theta_0}(X_i)\\
&&-(S_2+\theta_0\theta_0^\t)^{-1}(\Omega_0+\theta_0\theta_0^\t)\delta_\vartheta+ \alpha_n |\vartheta-\theta_0|  + o(n^{-1/2}) \label{paris}
\endy
almost surely, where
$\varpi_\theta(x)=E(X|X^\t\theta=x^\t\theta)-x,\ \alpha_n = o(1) $
uniformly in $\vartheta$ and
\beginn
&&\Omega_{n\vartheta}\stackrel{def}{=}\frac{1}{n}
\sum\limits_{j}b^2_j\mu_{\vartheta}(X_j)\{(\nu/\mu)_\vartheta(X_j)-X_j\}
 \times \{(\nu/\mu)_\vartheta(X_j)-X_j\}^\t\\
&&\Omega_0=E[
\{m'({X}^\t\theta_0)\}^2\mu_{\theta_0}(X)\{(\nu/\mu)_{\theta_0}(X)-X\}\{(\nu/\mu)_{\theta_0}(X)-X\}^\t
]
\endn
\end{Theorem}

\begin{Remark} \rm 
 In Lemma \ref{ste}, we prove that if $\delta_\vartheta\ne 0,$
\beginy
 0<
|(S_2+\theta_0\theta_0^\t)^{-1}(\Omega_0+\theta_0\theta_0^\t)\delta_\vartheta|/|\delta_\vartheta|<
1. \label{lilo}
\endy
This implies that  the effect on $\hat\theta-\theta_0$ of the initial estimate error $\vartheta-\theta_0$ 
  decreases geometrically.
\end{Remark}

\begin{Remark} \rm
 Theorem \ref{T0} is proved under the assumption that
$\{(X_i,Y_i)\}_{i=1}^\infty$ are I.I.D. observations. It is
possible, however, to extend this result for time series
observations provided that the time dependency (usually measured by
mixing coefficient) are weak enough. For example, the  stationary
$\beta-$ mixing  processes, which satisfies
\beginn
\beta(k)=\sup\limits_{A\in \mathcal F_{-\infty}^a,B\in\mathcal
F_{a+k}^\infty}|P(B)-P(B|A)|\to 0,\quad \mbox{ as }k\to \infty,
\endn
where $\mathcal F_{a}^b$ is the $\sigma-$algebra generated by
$\{(X_i,Y_i)\}_{i=a}^b$.
\end{Remark}
 %The rationality behind the above conjecture is that most of the lemmas which
%are used in the proof can be replaced by their counterparts in the
%time series setting, namely Bernstein's ineuality by Theorem 1.4 in Bosq
%(1998), and Lemma \ref{koro} by Theorem 2 in Sun et al (1997). The
%remaining issue is whether Lemma \ref{gin} is still true for mixing
%processes; that is, for a degenerate kernel $g(.)$ and mixing
%processes $(X_i,i=1,\cdots,n)$ with weak enough dependency
%structure, do we still have
%\beginn
%\lim \sup \limits_{n}\frac{1}{ n \log \log n} \sum\limits_{ i\ne
%j}g(X_i,X_j) < \infty \quad a.s.
%\endn
%Heuristically we do, a related fact being that the parallel
%asymptotic normality result is proved in Fan and Li (1999), i.e.
%\beginn
%n^{-1} \sum\limits_{ i\ne j}g(X_i,X_j) \to
%N(0,\sigma^2)\quad\mbox{for some constant }\sigma>0.
%\endn
%for strictly stationary $\beta-$mixing processes satisfying some
%regular conditions on $\beta(k)$.
%
\begin{Lemma}\label{LL} Under conditions in Theorem \ref{T0}, we have
\beginy
&(n^2h)^{-1}R_{n1}=\frac{1}{n}\sum\limits_{i}\varphi(\e_i)b_i\{\varpi
f\}_{\theta_0}(X_i)-\Omega_{n\vartheta}\delta_\vartheta+ \alpha_n
|\vartheta-\theta_0|  + o(n^{-1/2}) \ a.s. &\label{fff}
\endy
\end{Lemma}

\noindent\noindent{\bf Proof of Theorem \ref{T0}}. Based on (\ref{fff}), it
suffices to prove that
\beginy
&
\hat\theta-\theta_0=\{n^2h(S_{2}+\theta_0\theta_0^\t)\}^{-1}(R_{n1}-n^2h\theta_0\theta_0^\t\delta_\vartheta) \ a.s. &\label{Hu}
\endy
  As the first step to prove (\ref{Hu}), we show   in
Lemma \ref{wuhu} and Lemma \ref{anhui} that  for each fixed $\theta$,
\beginy
(n^2ha_{n\vartheta}^{2})^{-1}[R_{n2}(\theta)-ER_{n2}(\theta)]=o(1)\
a.s.\label{ricci1}
\endy
 This together with (\ref{rachel}) and the fact that
$G_{n\vartheta}=n^2hS_2\{1+O(\delta_{\vartheta})\}$ imply that for any fixed $\theta$,
\beginn
 (n^2ha_{n\vartheta}^{2})^{-1}[\tilde\Phi_n(\theta)-\delta_\theta^\t
(R_{n1}+\theta_0\theta_0^\t\delta_\vartheta)-\frac{1}{2}n^2h\delta_\theta^\t
(S_{2}+\theta_0\theta_0^\t)\delta_\theta]\to 0\ a.s.
\endn
As both $\tilde\Phi_n(\theta)-\delta_\theta^\t
(R_{n1}+\theta_0\theta_0^\t\delta_\vartheta)$ and $\delta_\theta^\t
(S_{2}+\theta_0\theta_0^\t)\delta_\theta$ are convex in $\theta$, it
follows from Lemma \ref{Pollard} that for any compact set
$\Theta_{n\theta}\subset
 \Theta_n$(convex open set),
\beginy
\sup\limits_{\theta\in
\Theta_{n\theta}}(n^2ha_{n\vartheta}^{2})^{-1}|\tilde\Phi_n(\theta)-\delta_\theta^\t
(R_{n1}+\theta_0\theta_0^\t\delta_\vartheta)-\frac{1}{2}n^2h\delta_\theta^\t
(S_{2}+\theta_0\theta_0^\t)\delta_\theta|\to 0\ a.s.\label{neurotic}
\endy
Let
$\eta_n=\{n^2h(S_{2}+\theta_0\theta_0^\t)\}^{-1}(R_{n1}+\theta_0\theta_0^\t\delta_\vartheta)$.
Now we are ready to prove the equivalent of (\ref{Hu}), i.e. with
probability $1$, for any $\delta>0$,
$|\hat\theta-\theta_0-\eta_n|/a_{n\vartheta}\le \delta$ for large
$n$.

 First note that as $\theta_0+\eta_n$ is bounded with
probability $1$,  $\Theta_n$ can be chosen to contain $B_n^\delta$, a
closed ball with center $\theta_0+\eta_n$ and radius
$a_{n\vartheta}\delta $. Replace $\Theta_{n\theta}$ in
(\ref{neurotic}) by $B_n^\delta$, we have
\beginy
\Delta_n\equiv\sup\limits_{\theta\in B_n^\delta
}(n^2ha_{n\vartheta}^{2})^{-1}|\tilde\Phi_n(\theta)-\delta_\theta^\t
(R_{n1}-\theta_0\theta_0^\t\delta_\vartheta)-\frac{1}{2}n^2h\delta_\theta^\t
(S_{2}+\theta_0\theta_0^\t)\delta_\theta|=o(1)\ a.s. \label{hua}
\endy
Now consider the behavior of $\tilde\Phi_n(\theta)$ outside
$B_n^\delta.$ Suppose $\theta=\theta_0+\eta_n+a_{n\vartheta}\beta
\nu$, for some $\beta>\delta$ and  $\nu$ a unit vector. Define
$\theta^*$ as the boundary point of $B_n^\delta$ that lies on the
line segment from $\theta_0+\eta_n$ to $\theta,$ i.e.
$\theta^*=\theta_0+\eta_n+ a_{n\vartheta} \delta\nu$. Convexity of
$\Phi_n(\theta)$ and the definition of $\Delta_n$ imply
\beginn
\frac{\delta}{\beta}\tilde\Phi_n(\theta)+(1-\frac{\delta}{\beta})\tilde\Phi_n(\theta_0+\eta_n)&\ge&
\tilde\Phi_n(\theta^*)\\
&\ge& \frac{1}{2}n^2h\delta^2a_{n\vartheta}^2\nu^\t
(S_{2}+\theta_0\theta_0^\t)\nu\\
&&-\frac{1}{2}(n^2h)^{-1}R_{n1}^\t
(S_{2}+\theta_0\theta_0^\t)^{-1}R_{n1}-n^2ha_{n\vartheta}^{2}\Delta_n
\\
&\ge& \frac{1}{2}n^2h\delta^2a_{n\vartheta}^2\nu^\t
(S_{2}+\theta_0\theta_0^\t)\nu+\tilde\Phi_n(\theta_0+\eta_n)-2n^2ha_{n\vartheta}^{2}\Delta_n.
\endn
It follows that
\beginn
\inf\limits_{|\theta-\theta_0-\eta_n|> \delta a_{n\vartheta}
}\tilde\Phi_n(\theta)\ge
\tilde\Phi_n(\theta_0+\eta_n)+\frac{\beta}{\delta}n^2ha_{n\vartheta}^2[\frac{1}{2}\delta^2\nu^\t
(S_{2}+\theta_0\theta_0^\t)\nu-2\Delta_n].
\endn
As $S_{2}+\theta_0\theta_0^\t$ is positive definite, then according
to (\ref{hua}), with probability $1$, $\delta^2\nu^\t
S_{2}\nu>4\Delta_n$ for large enough $n$. This implies that for any
$\delta>0$ and for large enough $n$, the minimum of
$\tilde\Phi_n(\theta)$ must occur within $B_n^\delta$. This implies
(\ref{Hu}).$\hspace{\fill}\blacksquare$

\section*{Appendix}
\noindent{\bf Proof of Lemma \ref{LL}}.
Write
\beginn
R_{n1}(\theta) =\sum\limits_{i,j}K^\vartheta_{ij}\varphi(\e_i) b_j
X_{ij}+\sum\limits_{i,j}K^\vartheta_{ij}\varphi(\e_i)(\hat b_j-b_j)
X_{ij}+\sum\limits_{i,j}K^\vartheta_{ij}\hat b_j
X_{ij}\{\varphi(Y_{ij})-\varphi(\e_i)\},
\endn
where $E_j$ denotes expectation taken w.r.t $X_j$ for given $X_i$.
We will show that
\beginy
\frac{1}{n^2h}\sum\limits_{i,j}K^\vartheta_{ij}\varphi(\e_i) b_j
X_{ij}=\frac{1}{n}\sum\limits_{i}\varphi(\e_i)b_i\{\varpi
f\}_{\theta_0}(X_i)+O\{(\log\log n/n)^{1/2}(h^2+\delta_\vartheta) \}, \label{rr}
\endy
 which together with Lemma \ref{chi} lead to (\ref{fff}).

First note that
\beginn
&& E_j[K^\vartheta_{ij} b_j X_{ij}/h]=b_i\{\varpi
f\}_{\vartheta}(X_i)-\delta_\vartheta m''(X_i^\t\theta_0)\{\Sigma
f\}_{\theta_0}(X_i)\\
&&\hspace{3cm}+h^2 b_i\{\varpi
f\}''_{\theta_0}(X_i)+O(|\delta_\vartheta|^2+h^4),
\endn
This together with
Lemma 7.8 in Xia and Tong (2006),
we have
\beginn
 && \frac{1}{n^2h}\sum\limits_{i,j}K^\vartheta_{ij}\varphi(\e_i) b_j
X_{ij}=\frac{1}{n}\sum\limits_{i}\varphi(\e_i)b_i\{\varpi
f\}_{\theta_0}(X_i)+O\{(\log\log n/n)^{1/2}(h^2+\delta_\vartheta) \},
\endn
from which follows (\ref{rr}), as $\{\varpi
f\}_{\vartheta}(.)$ is lipschitz continuous in $\vartheta$.
$\hspace{\fill}\blacksquare$
\begin{Lemma}\label{LL0}
\beginy
m_\vartheta(X_j)-a_j&=&b_j\delta_\vartheta^\t\{(\nu/\mu)_\vartheta(X_j)-X_j\}+o(|\delta_\vartheta|),\
\label{obama}\\
 m'_\vartheta(X_j)-b_j&=&b_j\delta_\vartheta^\t
\{(\mu\nu'-\mu'\nu)/\mu^2\}_\vartheta(
X_j)+o(|\delta_\vartheta|),\label{obama1}
\endy
\end{Lemma}
\noindent{\bf Proof} It follows from the
property of conditional expectation that
\beginn
 E\{\rho(Y-a)|X^\t\vartheta=x^\t\vartheta\}&=&E[E\{\rho(Y-a)|X\}|X^\t\vartheta=x^\t\vartheta]\\
 &=&E[G\{m(\theta_0^\t X)-a;X\}|X^\t\vartheta=x^\t\vartheta].
 \endn
Using the differentiability of $G(t;X)$ in $t$, we have
\beginn
G\{m(\theta_0^\t X)-a;X\}=G(0;X)+g(X)(m(\theta_0^\t
X)-a)^2/2+O\{(m(\theta_0^\t X)-a)^3\}.
\endn
If $X^\t\vartheta=x^\t\vartheta$ and $\delta_\vartheta=o(1)$,
 $m(\theta_0^\t X)-m(\theta_0^\t x)=O\{\theta_0^\t(X-x)\}=O\{\delta_\theta^\t(X-x)\}=o(1).$ Therefore for every $a$ near $m(\theta_0^\t X)$ (whence $m(\theta_0^\t x)\ $ [{\color{red} WHY}] ),
\beginn
E[G\{m(\theta_0^\t X)-a;
X\}|X^\t\vartheta=x^\t\vartheta]-E[G(0;X)|X^\t\vartheta=x^\t\vartheta]\hspace{4cm}\\
\rightarrow\frac{1}{2} E[g(X)(m(\theta_0^\t
X)-a)^2|X^\t\vartheta=x^\t\vartheta].
\endn
As $\rho(.)$ is convex, we can argue this convergence is in fact uniform over all $a$
near $m(\theta_0^\t X)$, which implies that the  minima of
$E[G\{m(\theta_0^\t X)-a;X\}|X^\t\vartheta=x^\t\vartheta]$ is also
approximately the minima of $E[g(X)(m(\theta_0^\t
X)-a)^2|X^\t\vartheta=x^\t\vartheta]$. We have
\beginn
m(\theta_0^\t X)=m(\theta_0^\t x)+m'(\theta_0^\t
x)\theta_0^\t(X-x)+C\{\theta_0^\t(X-x)\}^2,
\endn
\vspace{-1.3cm}
\beginy
\n E[g(X)(m(\theta_0^\t
X)-a)^2|X^\t\vartheta=x^\t\vartheta]&=&2m'(\theta_0^\t
x)\{m(\theta_0^\t
x)-a\}\delta_\vartheta^\t\{\nu_\vartheta(x^\t\vartheta)-x\mu_\vartheta(x^\t\vartheta)\}\\
&&+\{m(\theta_0^\t
x)-a\}^2\mu_\vartheta(x^\t\vartheta)+O(|\delta_\vartheta|^2).\label{snow}
\endy
Take derivative with respect to $a$ and (\ref{obama}) follows.

To prove (\ref{obama1}),  for any $t\to 0$, mimicking (\ref{snow}),
\beginn
&&E[g(X)\{m(\theta_0^\t X)-a\}^2|X^\t\vartheta=x^\t\vartheta+t]\\
&=&2m'(\theta_0^\t x)\{m(\theta_0^\t
x)-a\}
E[g(X)\{t+\delta_\vartheta^\t(X-x)\}|X^\t\vartheta=x^\t\vartheta+t]\\
&&+\{a-m(\theta_0^\t
x)\}^2\mu_\vartheta(x^\t\vartheta+t)+O(|\delta_\vartheta|^2)\\
&=&\{a-m(\theta_0^\t
x)\}^2\mu_\vartheta(x^\t\vartheta+t)+2tm'(\theta_0^\t
x)\{m(\theta_0^\t
x)-a\}\mu_\vartheta(x^\t\vartheta+t)+O(t^2|\delta_\vartheta|^2)\\
&& +2m'(\theta_0^\t x) \{m(\theta_0^\t
x)-a\}\delta_\vartheta^\t \{\nu_\vartheta(x^\t\vartheta+t)-x\mu_\vartheta(x^\t\vartheta+t)\}.
\endn
Again take derivative with respect to $a$ and by the definition of $m_\vartheta(.)$,
we have
\beginn m_\vartheta(\vartheta^\t x+t)\approx m(\theta_0^\t
x)+tm'(\theta_0^\t x)+m'(\theta_0^\t
x)\delta_\vartheta^\t\{(\nu/\mu)_\vartheta(x^\t\vartheta+t)-x\},
\endn
Recall that from (\ref{obama}),
$m_\vartheta(\vartheta^\t x)\approx m(\theta_0^\t x)+m'(\theta_0^\t
x)\delta_\vartheta^\t\{(\nu/\mu)_\vartheta(x^\t\vartheta)-x\}
+O(|\delta_\vartheta|^2).$ Subtract this from the equation above and
suppose the first order derivative of $\mu_\vartheta(.)$ and
$\nu_\vartheta(.)$ are both Lipschitz continuous, we have
\beginn
&&m_\vartheta(\vartheta^\t x+t)-m_\vartheta(\vartheta^\t
x)\\
&\approx &tm'(\theta_0^\t x)+m'(\theta_0^\t x)\delta_\vartheta^\t
\{(\nu/\mu)_\vartheta(x^\t\vartheta+t)-(\nu/\mu)_\vartheta(x^\t\vartheta)\}\\
&=&tm'(\theta_0^\t x)+tm'(\theta_0^\t
x)\delta_\vartheta^\t\{(\mu\nu'-\mu'\nu)/u^2\}_\vartheta(x^\t\vartheta)+O(t^2).
\endn
Divide this over $t$ and let $t\to 0$, we will have (\ref{obama1}).
$\hspace{\fill}\blacksquare$
\begin{Lemma}\label{great}$E_iK^\vartheta_{ij}\varphi(Y^*_{ij})=\frac{1}{2}m^{''}(X_j^\t\theta_0)(fg)
_\vartheta(X_j)h^3+O(h^4)+o(h\delta_\vartheta),\label{hgg}$
\beginy
\n E_i K^\vartheta_{ij}\varphi(Y^*_{ij})X_{ij}^\t\vartheta &=&
h^4\Big\{\frac{1}{2}m^{''}(X_j^\t\theta_0)(f\mu)'_\vartheta(X_j)
\\
&&+\frac{1}{6}m^{(3)}(X_j^\t\theta_0)(f\mu)_\vartheta(X_j)\Big\}+O(h^4\delta_\vartheta+h^6).\label{obama2}
\endy
\end{Lemma}
\noindent{\bf Proof} Based on
(\ref{obama}) and (\ref{obama1}), we have

\beginn
&&m(X_i^\t\theta_0)-m_\vartheta(X_j)-m'_\vartheta(X_j)X_{ij}^\t\vartheta\\
&=&m(X_i^\t\theta_0)
-m(X_j^\t\theta_0)-b_j\delta_\vartheta^\t\{(\nu/\mu)_\vartheta(X_j)-X_j\}\\
&&
 -\{b_j+b_j\delta_\vartheta^\t
\{(\mu\nu'-\mu'\nu)/\mu^2\}_\vartheta( X_j)\}X_{ij}^\t\vartheta+o(|\delta_\vartheta|)\\
&=&b_jX_{ij}^\t\delta_\vartheta+\frac{1}{2}m^{''}(X_j^\t\theta_0)(\theta_0^\t
X_{ij})^2+\frac{1}{6}m^{(3)}(X_j^\t\theta_0)(\theta_0^\t X_{ij})^3
\\
&&-b_j\delta_\vartheta^\t \{(\mu\nu'-\mu'\nu)/\mu^2\}_\vartheta(
X_j)X_{ij}^\t\vartheta-b_j\delta_\vartheta^\t\{(\nu/\mu)_\vartheta(X_j)-X_j\}\\
&&+o(|\delta_\vartheta|)+O\{(X_{ij}^\t\vartheta)^4+\delta_\vartheta\}.
\endn
As $m(X_i^\t\theta_0)-m_\vartheta(X_j)-m'_\vartheta(X_j)X_{ij}^\t\vartheta=o(1)$, by the continuity of $G_1(t;X)$ in $t$, we have
\beginy
\n
&&E[\varphi\{Y_{i}-m_\vartheta(X_j)-m'_\vartheta(X_j)X_{ij}^\t\vartheta\}|X_i]\\
\n &=&G_1\{m(X_i^\t\theta_0)-m_\vartheta(X_j)-m'_\vartheta(X_j)X_{ij}^\t\vartheta;X_i\}\\
\n&=&b_j\delta_\vartheta^\t
g(X_i)X_{ij}-b_j\delta_\vartheta^\t\{(\nu/\mu)_\vartheta(X_j)-X_j\}g(X_i)-b_j\delta_\vartheta^\t
\{(\mu\nu'-\mu'\nu)/\mu^2\}_\vartheta( X_j)g(X_i)X_{ij}^\t\vartheta
\\
&&+\frac{1}{2}m^{''}(X_j^\t\theta_0)g(X_i)(\theta_0^\t
X_{ij})^2+\frac{1}{6}m^{(3)}(X_j^\t\theta_0)g(X_i)(\theta_0^\t
X_{ij})^3+o(|\delta_\vartheta|)+O((X_{ij}^\t\vartheta)^4),\label{sh}
\endy
and thus
\beginn
E_i[K^\vartheta_{ij}\varphi\{Y_{i}-m_\vartheta(X_j)-m'_\vartheta(X_j)X_{ij}^\t\vartheta\}]
=\frac{1}{2}m^{''}(X_j^\t\theta_0)(g
f)_\vartheta(X_j)h^3+o(h|\delta_\vartheta|)+O(h^4).
\endn
 Similarly (\ref{obama2})
 follows from (\ref{sh}) and the following
facts
\beginn
E[g(X_i)X_{ij}|X_i^\t\vartheta=X_j^\t\vartheta+hu]&=&\nu_\vartheta(X_j^\t\vartheta+hu)-X_j\mu_\vartheta(X_j^\t\vartheta+hu)\\
&=&\nu_\vartheta(X_j^\t\vartheta)+hu\nu'_\vartheta(X_j^\t\vartheta)
-X_j\mu_\vartheta(X_j^\t\vartheta)\\
&& -huX_j\mu'_\vartheta(X_j^\t\vartheta)+O(h^2),\\
E[g(X_i)|X_i^\t\vartheta=X_j^\t\vartheta+hu]&=&\mu_\vartheta(X_j^\t\vartheta)
+hu\mu'_\vartheta(X_j^\t\vartheta)+O(h^2),  \\
\int K(u)E[g(X_i)X_{ij}|X_i^\t\vartheta=X_j^\t\vartheta+hu]hudu\hspace{-.2cm}&=\hspace{-.2cm}&
h^2\{(f\nu')_\vartheta(X_j^\t\vartheta)-X_j(f\mu')_\vartheta(X_j^\t\vartheta)\}\\
&&+h^2\{(f'\nu)_\vartheta(X_j^\t\vartheta)-X_j(f'\mu)_\vartheta(X_j^\t\vartheta)\}+O(h^4),\\
\int K(u)E[g(X_i)|X_i^\t\vartheta=X_j^\t\vartheta+hu]hudu&=&h^2(\mu'f+\mu
f')_\vartheta(X_j^\t\vartheta)+O(h^4),\\
\int K(u)E[g(X_i)|X_i^\t\vartheta=X_j^\t\vartheta+hu]h^2u^2du &=& h^2(\mu
f)_\vartheta(X_j^\t\vartheta)+O(h^4).\hspace{3.5cm}\blacksquare
\endn

\begin{Lemma} \label{Pollard} Let $\{\lambda_n(\theta):\theta\in \Theta\}$ be a
sequence  of random convex functions defined on a convex, open
subset $\Theta$ of $R^d$. Suppose $\lambda(\theta)$ is a real valued
function on $\Theta$ such that $\lambda_n(\theta)$ tends to
$\lambda(\theta)$ for each $\theta$ almost surely, Then for each
compact set $K$ of $\Theta$,  with probability $1$,
\beginn
\sup\limits_{\theta\in K } |\lambda_n(\theta) - \lambda(\theta)|\to
0.
\endn
\end{Lemma}
\noindent{\bf Proof}  The condition can be restated as follows: for
any fixed $\theta\in \Theta$, there exists some
$\Omega_\theta\subseteq \Omega$, such that $P(\Omega_\theta)=1$ and
\beginn
\lambda_n(\omega,\theta) - \lambda(\theta)\to 0,\mbox{ for any
}\omega\in \Omega_\theta.
\endn
The conclusion can be restated that for each compact set $K$ of
$\Theta$, there exists some $\Omega_0\subseteq \Omega$, such that
\beginn
P(\Omega_0)=1\quad \mbox{and  }\sup\limits_{\theta\in K }
|\lambda_n(\omega,\theta) - \lambda(\theta)|\to 0,\mbox{ for any
}\omega\in \Omega_0.
\endn
For such uniformity of the convergence, it is enough to
consider the case where $K$ is a cube with edges parallel to the
coordinate directions $e_l,\cdots ,e_d$. Every compact subset of
$\Theta$ can be covered by finitely many such cubes. Let $\Im_0\equiv K$ and $K^{+\delta_0}$ be the larger cube
constructed by adding an extra layer of cubes with sides $\delta_0$
to $K$. Suppose $\delta_0>0$ is small enough such that
$K^{+\delta_0}\subset \Theta.$ Define $\mho_0$ for the finite set of
all vertices of all the cubes that make up $K^{+\delta^0}$.

Now for $k=1,2,\cdots $, let $\epsilon_k =k^{-1}$. As
convexity implies continuity, there is a $0<\delta^k<\delta^{k-1}$
such that $\lambda(.)$ varies by less than $\epsilon_k/(d+1)$ over
each cube of side $3\delta^k$ that intersects $K$.  Partition each
cube in $\Im_{k-1}$ into a union of cubes with side at most
$\delta^k$ and denote by $\Im_{k}$ the resulted union of cubes. Then
expand $K$ to a larger cube $K^{+\delta^k}$ by adding an extra layer
of these $\delta^k-$cubes around each face. As
$\delta^k<\delta^{k-1}$, $K^{+\delta^k}\subset K^{+\delta^{k-1}}$ is
still within $\Theta$. Define
\beginn
\mho_k &=&\{\mbox{ vertices of all the }\delta^k-\mbox{cubes that
make up
}K^{+\delta^k}\}\bigcup\mho_{k-1}\\
&\equiv&\{\mbox{ vertices of all the }\delta^k-\mbox{cubes that make
up }K^{+\delta^k}\}\bigcup \{\mho_{k-1}\bigcap K^c\} \endn
 and
\beginn
\Omega_k=\bigcap\limits_{\theta\in \mho_k}\Omega_\theta.
\endn
As $\mho_k$ is finite, we have $P(\Omega_k)=1$ and
\beginy
\mbox{for any }\omega\in \Omega_k,\quad
M_n^k(\omega)=\sup\limits_{\theta\in
\mho_k}|\lambda_n(\omega,\theta) - \lambda(\theta)|\to 0.\label{ha}
\endy

\noindent  We first establish the connection between $M_n^k(\omega)$
and the upper bound for $\lambda_n(\omega,\theta)-\lambda(\theta),$
over $ \theta\in K,$ for any given $\omega\in \Omega_k$.

\noindent  For any fixed $k=1,2,\cdots$, each $\theta$ in $K$ lies
within a $\delta^k$-cube with vertices $\{\theta_i\}\in \mho_k$; it
can be written as a convex combination $\sum_i\alpha_i\theta_i$ of
those vertices, i.e.
\beginn
\theta=\sum_{\theta_i\in \mho_k}\alpha_{i}\theta_i, \quad
\sum_{\theta_i\in \mho_k}\alpha_{i}=1.
\endn
Then for any given $\omega\in \Omega_k$, convexity of
$\lambda_n(\omega,\theta)$ in $\theta$ gives
\beginn
\n\lambda_n(\omega,\theta)&\le &\sum_{\theta_i\in \mho_k}\alpha_{i}\lambda_n(\omega,\theta_i)\\
\n&= &\sum_{\theta_i\in
\mho_k}\alpha_{i}\{\lambda_n(\omega,\theta_i)-\lambda(\theta_i)\}+
\sum_{\theta_i\in \mho_k}\alpha_{i}\{\lambda(\theta_i)-\lambda(\theta)\}+\lambda(\theta)\\
\n &\le & M_n^k(\omega)+\max\limits_{\theta_i\in
\mho_k}|\lambda(\theta_i) - \lambda(\theta)|+\lambda(\theta).
\endn
Therefore,
\beginy
\lambda_n(\omega,\theta)-\lambda(\theta)\le
M_n^k(\omega)+\epsilon_k.\label{lone}
\endy

\noindent Next we establish the companion lower bound.  For any
fixed $k=1,\cdots$, each $\theta$ in $K$ lies within a
$\delta^k$-cube with a vertex $\theta_0$ in $K\bigcap \mho_k$:
 \beginn
 \theta=\theta_0+\sum\limits_{i=1}^d\delta_ie_i,\ \mbox{with
 }|\delta_i|\le\delta^k,\ i=1,\cdots,d.
\endn
 Without loss of generality,  suppose $\delta_i\ge 0$ for each
 $i=1,\cdots,d.$ Define
 \beginn
\theta_{ik}=\theta_0-\delta_i'e_i,\ \quad\mbox{where
 }\delta_i'\equiv \min\{c\ge \delta_k:\theta_0-ce_i\in \mho_k\},
 \ i=1,\cdots,d
 \endn
Note that as $\theta_0\in K\bigcap \mho_k$, $\delta_i'$ must exist
and $\delta_i'<2\delta^k$, for all $i=1,\cdots,d.$

\noindent Write $\theta_0$  as a convex combination of $\theta$ and
these $\theta_{ik}$:
\beginn
\theta_0=\frac{\prod_{j=1}^d \delta_j' }{\prod_{j=1}^d
\delta_j'+\sum_{j=1}^d\delta_j\prod_{l\ne
j}\delta_l'}\theta+\sum\limits_{i=1}^d\frac{\delta_i\prod_{j\ne i}
\delta_j' }{\prod_{j=1}^d \delta_j'+\sum_{j=1}^d\delta_j\prod_{l\ne
j}\delta_l'}\theta_{ik}.
\endn
Denote these convex weights by $\beta$ and $\{\beta_i\}$. As
$\delta_j\le \delta^k\le \delta_j'$, we have $\beta\ge 1/(d+1)$ and
 \beginn
\beta\lambda_n(\omega,\theta)&\ge&
\lambda_n(\omega,\theta_0)-\sum\limits_i\beta_i\lambda_n(\omega,\theta_{ik})\quad(\mbox{
convexity of }\lambda_n(\omega,\theta)\mbox{ in }\theta)\\
&\ge&\lambda(\theta_0)-\sum\limits_i\beta_i\lambda(\theta_{ik})-
2M_n^k(\omega)\quad(\mbox{ from }(\ref{ha}))\\
&\ge&\lambda(\theta)-\epsilon_k/(d+1)-\sum\limits_i\beta_i[\lambda(\theta)+\epsilon_k/(d+1)]-2M_n^k(\omega)\\
&=& \beta\lambda(\theta)-2\epsilon_k/(d+1)-2M_n^k(\omega)
\endn
 where the third inequality is due to the definition of $\delta^k$
 and the fact that there exists a cube of side $3\delta^k$ which
 contains both
  $\theta_{ik}$ and $\theta_0$. As $\beta\ge 1/(d+1)$,
\beginn
\lambda_n(\omega,\theta)-\lambda(\theta)\ge
-2\epsilon_k-2(d+1)M_n^k(\omega).
\endn
 This together with
(\ref{lone}) implies that for any $k=1,2,\cdots $, there exists some
$\Omega_k(\supseteq \Omega_{k+1})$ such that $P(\Omega_k)=1$ and
\beginn
\forall \omega\in \Omega_k,\ \sup\limits_{\theta\in
K}|\lambda_n(\omega,\theta)-\lambda(\theta)|\le (d+1)
M_n^k(\omega)+2k^{-1}.
\endn
Let $\Omega_0\equiv \bigcap_{k=1}^\infty \Omega_k$. As $\Omega_k$ is
a decreasing sequence and $P(\Omega_k)=1$, we have $P(\Omega_0)=1$
and for any $\omega\in \Omega_0$,
\beginy
 \sup\limits_{\theta\in
K}|\lambda_n(\omega,\theta)-\lambda(\theta)|\le
(d+1)M_n^k(\omega)+2k^{-1}, \mbox{ for all }k\ge 1.\label{rt}
\endy
Note that as $n\to \infty,$ $M_n^k(\omega)\to 0$ for each fixed $k$,
as in (\ref{ha}). Take limit of both sides of (\ref{rt})
\beginn
\lim_{n\to \infty} \sup\limits_{\theta\in
K}|\lambda_n(\omega,\theta)-\lambda(\theta)|\le \lim_{n\to
\infty}M_n^k(\omega)+k^{-1}=k^{-1},\mbox{ for all }k\ge 1.
\endn
This is equivalent to that with probability $1,$ $ \lim_{n\to
\infty} \sup\limits_{\theta\in
K}|\lambda_n(\omega,\theta)-\lambda(\theta)|\to 0. $
$\hspace{\fill}\blacksquare$ \vspace{.5cm}

We now list a number of facts in the literature that will be used in
our proofs later.
%\begin{Lemma}\label{Bern1}\rm [Bernstein's inequality]
%Let $X_1,\cdots,X_n$ be
%independent zero-mean real valued random variables and there exists
%$c>0$ such that the following Cramer's condition are satisfied
%\beginn
%E|X_i|^k\le c^{k-2}k!EX_i^2<+\infty,\ i=1,\cdots,n;\ k=3,4,\cdots\
%\endn
% Let $S_n=\sum_{i=1}^n X_i,$ then
% \beginn
%P(|S_n|\ge t)\le 2\exp\Big(-\frac{t^2}{4\sum_{i=1}^n
%EX_i^2+2ct}\Big),\ t>0.
% \endn
%\end{Lemma}

%\begin{Lemma}\label{gin}
%\rm [Theorem 1.1. Gin\'{e} et al] Let $X,\ Y,\ X_i,\ i=1,\cdots,$
% be i.i.d. random variables taking
%values in $S$ and let $g: S\times S \to R$ be a measurable function
%of two variables. Then,
%\beginn
%\lim \sup \limits_{n}\frac{1}{ n \log \log n} \sum\limits_{ i\ne
%j}g(X_i,X_j) < \infty \quad a.s.
%\endn
%if an only if the following three conditions hold:
%\begin{description}
%\item{ (a)} $g(.)$ is integrable and $Eg(X,y)=0$ for almost all $y\in
%S.$
%\item (b) There exists some $C < \infty$ such that    $E\{g^2(X,Y) \wedge u\} \le C\log \log
% u$ for all $u \ge 10$
%\item  (c) There exists some $C < \infty$ such that
%$\sup\{Eg(X,Y)f_1(X)f_2(Y) ;Ef_1^2(X)\le 1,\ Ef^2_2(X)\le 1,\ \|f_1
%\|_\infty < \infty,\ \|f_2 \|_\infty < \infty\} \le C$
%\end{description}
%\end{Lemma}

\begin{Lemma}\label{koro}\rm [Korolyuk et al, 1989]
Let $X_1, X_2,\cdots, X_n$ be i.i.d. random variables. With a
symmetric kernel $\Phi: X^ m \to R$, we consider the U-statistic
\beginn
U_n={n\choose m} \sum\limits_{l\le i_1 <\cdots<i_m\le
n}\Phi(X_{i_1},\cdots,X_{i_m})
\endn
 Let $\theta =
E\Phi(X_{1},\cdots,X_{m})<\infty$ and  for $c = 0, 1,\cdots,m,$
define
\beginn
&&\Phi_c (x_1,\cdots,x_c) = E (\Phi (X_1,\cdots,X_m )| X_1 =
x_1,\cdots, X_c = x_c),\quad, \Phi_0=\theta,\ \Phi_m=\Phi\\
&&g_c (x_1,\cdots,x_c) = \sum\limits_{d=0}
^c(-1)^{c-d}\sum\sum\limits_{l\le j_1 <\cdots<j_d\le c}\Phi_d
(x_{j1},\cdots,x_{jd} ),\quad \sigma_1^2=Eg^2_1(X_1)
\endn
Suppose $\sigma_1^2>0$ and for all $c= 1 ,\cdots, m$, $
Eg_c^{2c/(2c-1)}<\infty. $ The with probability $1$,
 \beginn
\lim \sup\limits_{n\to \infty}\frac{n^{1/2}(U_n-\theta)}
{(2m^2\sigma_1^2\log\log  n)^{ 1/2}} = 1\quad \blacksquare
\endn
\end{Lemma}

\begin{Lemma}\label{mao}\rm [Berbee's Lemma]  Let $(X,Y)$ be a
$R^d\times R^{d'}-$valued random vector. Then there exists a
$R^{d'}-$valued random vector $Y^*$ which has the same distribution
as $Y$ and
\beginy
Y^* \mbox{ is independent of }X;\ P(Y^*\ne
Y)=\beta(\sigma(X),\sigma(Y))\label{yang}
\endy
 where $\sigma(X)$ and $\sigma(Y)$) are the $\sigma-$algebra generated by
 $X$ and $Y$ respectively, and
 \beginn
 \beta[\sigma(X),\sigma(Y)]=E\sup_{A\in \sigma(Y) }
 |P( A)-P( A|\sigma(X))|\label{tax}
 \endn
\end{Lemma}

\begin{Lemma}\label{cher}
$\ \beta[\sigma(X_1,Y_1),\sigma(\hat a_j,\hat b_j )]=O\{(nh/\log^3
n)^{-1/4}\} $
\end{Lemma}
\noindent{\bf Proof} By the definition,
\beginn
 \beta[\sigma(X_1,Y_1),\sigma(\hat a_j,\hat b_j )]=E\sup_{A\in \sigma(\hat a_j,\hat b_j)}
 |P( A)-P( A|\sigma(X_1,Y_1))|
 \endn

According to results in Welsh (1996), $[(\hat a_j-E\hat a_j)/\sigma_1,(\hat b_j-E\hat
b_j)/\sigma_2]$ are asymptotically normal, where $
\sigma_1\equiv\{\Var\hat a_j\}^{1/2}=O\{(nh)^{-1/2}\}$ and
$\sigma_2\equiv\{\Var\hat b_j\}^{1/2}=O\{(nh^3)^{-1/2}\}$. Let
$\tau_n=(nh/\log n)^{-3/4}$ and rewrite
(\ref{lu1})
 as
\begin{equation*}
\hat a_j=E\hat a_j+\frac{1}{nh}\sum\limits_{i=2}^nK_{ij}^{\vartheta}
\varphi_{ij}+\frac{1}{nh}K_{1j}\varphi_1(X_1,Y_1)+O(\tau_n),
\end{equation*}
\begin{equation}
 \hat b_j=E\hat b_j+\frac{1}{nh^2}\sum\limits_{i=2}^n
\tilde\varphi_{ij}+\frac{1}{nh^2}\tilde\varphi_{1j}+O\{\tau_n/h\}.\label{exb}
\end{equation}
Note that $\varphi_{ij},\ \tilde\varphi_{ij},\ i=1,\cdots,n$ are
two sequences of zero-mean i.i.d. bounded random variables defined in (\ref{lu2}), whence
\beginn
  P\{\hat a_j\le t_1,\hat b_j\le t_2|Y_1,X_1\}
 &\le&
  P[\hat  a_j\le C\tau_n+t_1,
  \hat  b_j\le C\tau_n/h+t_2]\\
&\le& P\Big[(\hat a_j-E\hat a_j)/\sigma_1\le
(t_1-E\hat a_j+C\tau_n)/\sigma_1,\\
&&
(\hat b_j-E\hat b_j)/\sigma_2\le (t_2-E\hat b_j+C\tau_n/h)/\sigma_2\Big]\\
&=& P[\hat a_j\le t_1,\hat b_j\le t_2]+C(nh)^{1/2}\tau_n,
\endn
\vspace{-1cm}
\beginn
 P\{\hat a_j\ge t_1,\hat b_j\ge t_2|Y_1,X_1\}
 &\ge&
  P[\hat  a_j\ge t_1-C\tau_n ,
  \hat  b_j\ge t_2-C\tau_n/h]\\
&\ge& P\Big[(\hat a_j-E\hat a_j)/\sigma_1\ge
(t_1-E\hat a_j-C\tau_n)/\sigma_1,\\
&&
(\hat b_j-E\hat b_j)/\sigma_2\ge (t_2-E\hat b_j-C\tau_n/h)/\sigma_2\Big]\\
&=& P[\hat a_j\ge t_1,\hat b_j\ge t_2]-C(nh)^{1/2}\tau_n.
\endn
Therefore, \beginn |P\{\hat a_j\le t_1,\hat b_j\le
t_2|Y_1,X_1 \}-P\{\hat a_j\le t_1,\hat b_j\le t_2\}|\le
C(nh)^{-1/2}\tau_n=O\{(nh/\log^3 n)^{-1/4}\} .
\endn

\begin{Lemma}\label{gg} Under the assumptions (A1)--(A5), we have
\beginn
E\Phi_n(\theta)=\delta_\theta^\t E R_{n1}(\theta)+\delta_\theta^\t
G_{n\vartheta}\delta_\theta+o(n^2h|\delta_\theta|^2).
 \endn

\end{Lemma}
\noindent{\bf Proof} Apparently it suffices to show that
\beginn
&&EK^\vartheta_{ij}\{\rho(Y_1-\hat a_j-\hat b_j\theta^\t
X_{1j})-\rho(Y_1-\hat a_j-\hat b_j\theta_0^\t X_{1j})\}\\
 &=&\delta_\theta^\t E[K^\vartheta_{ij}\varphi(Y_1-\hat a_j-\hat
b_j\theta_0^\t X_{1j}) \hat b_jX_{1j}]+\delta_\theta^\t
E[K^\vartheta_{ij}X_{1j}X_{1j}^\t g(X_1)
\hat b^2_j]\delta_\theta+o(|\delta_\theta|^2).
\endn
By the continuity of $E[\rho(Y_1-\hat a_j-t \hat b_j)|\mathcal{X}]$ in $t$,
where $\mathcal{X}=\sigma(X_1,\cdots,X_n)$,
we have
\beginn
&&E\{\rho(Y_1-\hat a_j-\hat b_j\theta^\t
X_{1j})-\rho(Y_1-\hat a_j-\hat b_j\theta_0^\t
X_{1j})|\mathcal{X}\}\\
\n &=&\delta_\theta^\t X_{1j}E[\varphi(Y_1-\hat a_j-\hat
b_j\theta_0^\t X_{1j}) \hat b_j|\mathcal{X}] +\delta_\theta^\t X_{1j}X_{1j}^\t\delta_\theta
\partial
[E\{\varphi(Y_1-\hat a_j-\hat b_j t)\hat b_j|\mathcal{X}\}]/\partial t|_{t=X_{1j}^\t\theta_0} \\
\n &&+\delta_\theta^\t X_{1j}X_{1j}^\t\delta_\theta
\Big[\partial
[E\{\varphi(Y_1-\hat a_j-\hat b_j t)\hat b_j|\mathcal{X}\}]/\partial t|_{t=X_{1j}^\t\theta_0}
-\partial
[E\{\varphi(Y_1-\hat a_j-\hat b_j t)\hat b_j|\mathcal{X}\}]/\partial t|_{t=t^*}\Big]
\endn
where $t^*$ is some value between $\theta^\t X_{1j}$ and
$\theta_0^\t X_{1j}$. Taking expectations of both sides, we have
\beginy
&&EK^\vartheta_{ij}\{\rho(Y_1-\hat a_j-\hat b_j\theta^\t
X_{1j})-\rho(Y_1-\hat a_j-\hat b_j\theta_0^\t
X_{1j})\}\label{fei}\\
\n &=&\delta_\theta^\t E[K^\vartheta_{ij}\varphi(Y_1-\hat a_j-\hat
b_j\theta_0^\t X_{1j}) \hat b_jX_{1j}] +\delta_\theta^\t(\Delta_1+\Delta_2)\delta_\theta\\
\n \Delta_1&=&
E\{K^\vartheta_{ij}X_{1j}X_{1j}^\t\partial
[E\{\varphi(Y_1-\hat a_j-\hat b_j t)\hat b_j|\mathcal{X}\}]/\partial t|_{t=X_{1j}^\t\theta_0}\} \\
\n \Delta_2&=&
E\{K^\vartheta_{ij}X_{1j}X_{1j}^\t\partial
[E\{\varphi(Y_1-\hat a_j-\hat b_j t)\hat b_j|\mathcal{X}\}]/\partial t|_{t=X_{1j}^\t\theta_0}\}-\Delta_1
\endy
where $t^*$ is some value between $\theta^\t X_{1j}$ and
$\theta_0^\t X_{1j}$.

To study $\Delta_1$, we need to compute $\partial
[E\{\varphi(Y_1-\hat a_j-\hat b_j t)\hat b_j|\mathcal{X}\}]/\partial t$.
To this end, we apply Lemma \ref{mao} and Lemma \ref{cher}. Suppose $[\tilde a_j,\tilde b_j]$  has the
same distribution as $[\hat a_j,\hat b_j]$, but is independent of
$(Y_1,X_1)$ and $P([\tilde a_j,\tilde b_j]\ne [\hat a_j,\hat
b_j])=O\{(nh/\log^3 n)^{-1/4}\} $. Thus for any $\delta\to 0,$
\beginy
\n &&E[\varphi(Y_1-\hat a_j-\hat b_j(t+\delta )) \hat
b_j|\mathcal{X}]-E[\varphi(Y_1-\hat a_j-\hat b_jt) \hat
b_j|\mathcal{X}]\\
\n &=& E[\varphi\{Y_1-\tilde a_j-\tilde b_j(t+\delta)\}\tilde
b_j]-E[\varphi(Y_1-\tilde a_j-\tilde
b_jt)\tilde b_j|\mathcal{X}]\\
\n &&+ E[\{\varphi(Y_1-\hat a_j-\hat b_j(t+\delta))-\varphi(Y_1-\hat
a_j-\hat b_jt)\}\hat b_jI\{[\tilde a_j,\tilde
b_j]\ne [\hat a_j,\hat b_j]\}|\mathcal{X}]\\
\n &&- E[\{\varphi(Y_1-\tilde a_j-\tilde
b_j(t+\delta))-\varphi(Y_1-\tilde a_j-\tilde b_jt)\}\tilde
b_jI\{[\tilde
a_j,\tilde b_j]\ne [\hat a_j,\hat b_j]\}|\mathcal{X}]\\
&\equiv&\T_1+\T_2+\T_3\label{win}
\endy
Based on the definition of  $G_1(s;X)$, since  $Y_1$ is independent of $[\tilde
a_j,\tilde b_j]$,  we have
\beginy
 \n \T_1 &=&E[\{G_1(a_1-\tilde a_j-\tilde b_j(t+\delta);X_1)-G_1(a_1-\tilde
a_j-\tilde b_jt;X_1)\}\tilde b_j|\mathcal{X}]\\
&=&\delta E[G_2(a_1-\tilde a_j-\tilde b_jt;X_1)\tilde
b^2_j|\mathcal{X}]+o(\delta),\label{win3}
\endy
where the last equality follows from the continuity of $G_1(t;X)$ in $t$.

Next, we show that $\T_2=o(\delta)$.
 As we mentioned in the proof of Lemma \ref{cher}, $[v_1,v_2]\equiv[(\hat a_j-E\hat a_j)/\sigma_1,(\hat b_j-E\hat
b_j)/\sigma_2]$ are asymptotically normal, where
\beginn
\sigma_1\equiv\{\Var\hat a_j\}^{1/2}=O\{(nh)^{-1/2}\},\
\sigma_2\equiv\{\Var\hat b_j\}^{1/2}=O\{(nh^3)^{-1/2}\}.
\endn
Similarly construct $[\tilde v_1,\tilde v_2]$ from $\tilde a_j$ and
$\tilde b_j.$ Without loss of generality,
consider a small $\delta(>0).$ It is easy to understand that the conditional probability density function of $Y_1$
given $[v_1,v_2]$ is uniformly bounded. Therefore, for any given values of $\hat a_j$ and $\hat
b_j$ (equivalently $v_1$ and $v_2$),
\beginn
|E\{\varphi(Y_i-\hat a_j-\hat b_j(t+\delta))-\varphi(Y_i-\hat a_j-\hat
b_jt)|v_1,v_2\}|\le C\delta |\hat b_j|.
\endn
 Let $f(\tilde v_1,\tilde v_2|v_1,v_2)$ be the conditional probability density function
 of $(\tilde v_1,\tilde v_2)$ given $(v_1,v_2)$, and
 \beginn
&&g(v_1,v_2)=\int\limits_{[\tilde v_1,\tilde v_2]\ne
[v_1,v_2]}f(\tilde v_1,\tilde v_2|v_1,v_2)d\tilde v_1d\tilde
v_2.
\endn
As $\int f(v_1,v_2)g(v_1,v_2)dv_1dv_2=P([\tilde a_j,\tilde b_j]\ne [\hat a_j,\hat
b_j])=O\{(nh/\log^3 n)^{-1/4}\} $, we have
\beginn
|\T_2|\le C\delta\int \hat
 |b_j|f(v_1,v_2)g(v_1,v_2)dt_1dt_2=o(\delta).
 \endn

Similarly we can show that $\T_3=o(\delta)$. This together with
(\ref{win}) and (\ref{win3}) yields
\beginy
\partial
[E\varphi(Y_i-\hat a_j-\hat b_j t)\hat b_j|\mathcal{X}]/\partial
t=E[G_2(a_1-\tilde a_j-\tilde b_jt;X_1)\tilde
b^2_j|\mathcal{X}].\label{fei1}
\endy
Apply this result to $\Delta_1$ and $\Delta_2$, we have
\beginn
\Delta_1=E[K^\vartheta_{ij}X_{1j}X_{1j}^\t G_2(a_1-\tilde a_j-\tilde b_jX_{1j}^\t\theta_0;X_1)\tilde
b^2_j],\quad \Delta_2=O(\delta_\theta).
\endn
Plugging this into (\ref{fei})  leads to
\beginn
&&EK^\vartheta_{ij}\{\rho(Y_1-\hat a_j-\hat b_j\theta^\t
{X}_{1j})-\rho(Y_1-\hat a_j-\hat b_j\theta_0^\t
{X}_{1j})\}\\
\n &=&\delta_\theta^\t E[K^\vartheta_{ij}\varphi(Y_1-\hat a_j-\hat
b_j{X}_{1j}^\t\theta_0 ) \hat b_j{X}_{1j}]+\delta_\theta^\t
E[K^\vartheta_{ij}{X}_{1j}{X}_{1j}^\t G_2(a_1-\tilde a_j-\tilde
b_jX_{1j}^\t\theta_0;X_1)\tilde
b^2_j]\delta_\theta+o(|\delta_\theta|^2)\\
 &=&\delta_\theta^\t E[K^\vartheta_{ij}\varphi(Y_1-\hat a_j-\hat
b_j\theta_0^\t {X}_{1j}) \hat b_j{X}_{1j}]+\delta_\theta^\t
E[K^\vartheta_{ij}{X}_{1j}{X}_{1j}^\t g( X_1)
b^2_j]\delta_\theta+o(|\delta_\theta|^2)
\endn
where  the last equality follows from the continuity of
$G_2(t;X_1)$ in $t$ and (\ref{welsh}). $\hspace{\fill}\blacksquare$

\begin{Lemma}\label{chi}
Define $Z_{ij}=K^\vartheta_{ij}\hat b_j
X_{ij}\{\varphi(Y_{ij})-\varphi(\e_i)\}.$ Then
\beginy
&\hspace{-0.5cm}h^{-1}E_iZ_{ij}=-\delta_\vartheta^\t
b^2_j\{(\nu/\mu)_\vartheta(X_j)-X_j\}\{\nu_\vartheta(X_j)-X_j\mu_\vartheta(X_j)\}^\t
+o(|\delta_\vartheta|+n^{-1/2}),&\label{f1}\\
&\sum\limits_{i,j}(Z_{ij}-E_iZ_{ij})=o(n^2h\delta_\vartheta),&\label{f2}\\
&(nh)^{-1}\sum\limits_{i}K^\vartheta_{ij}\varphi(\e_i)(\hat
b_j-b_j) X_{ij}=o(n^{-1/2})+O\{\delta_\vartheta(nh/\log
n)^{-1/2}\} &\label{garner}
\endy
 uniformly in $\vartheta$.
\end{Lemma}
\noindent{\bf Proof} Once again we apply Lemma \ref{mao} and suppose $[\tilde a_j,\tilde
b_j]$ has the same distribution as $[\hat a_j,\hat b_j]$ and is
independent of $(X_1,Y_1)$. By Lemma \ref{cher}, $P([\tilde
a_j,\tilde b_j]\ne [\hat a_j,\hat b_j]\})=O\{(nh/\log^3 n)^{-1/4}\}$. Recall $
\mathcal{X} = \sigma(X_1, \cdots, X_n) $. Note that $E_1Z_{1j}= E[K_{1j}X_{1j}(\T_1-\T_2+\T_3)]$, where
\beginn
&&E[\{\varphi(Y_1-\hat a_j-X_{1j}^\t\theta_0\hat b_j)-\varphi(\e_1)\}\hat b_j|\mathcal{X}]=\T_1-\T_2+\T_3,\\
 && \T_1=E[\{\varphi(Y_1-\tilde a_j-\tilde b_jX_{1j}^\t\theta_0)-\varphi(\e_1)\}\tilde
b_j|\mathcal{X}]\\
 && \T_2= E[\{\varphi(Y_1-\tilde a_j-\tilde
b_jX_{1j}^\t\theta_0)-\varphi(\e_1)\}\tilde b_jI\{[\tilde a_j,\tilde b_j]\ne [\hat
a_j,\hat
b_j]\}|\mathcal{X}]\\
\n && \T_3=E[\{\varphi(Y_1-\hat a_j-\hat b_jX_{1j}^\t\theta_0)-\varphi(\e_1)\}\hat
b_jI\{[\tilde a_j,\tilde b_j]\ne [\hat a_j,\hat b_j]\}|\mathcal{X}].
\endn
Similar to (\ref{win3}),  we can conclude that
\beginy
\T_1&=&E[\{G_1(a_1-\tilde a_j-\tilde b_jX_{1j}^\t\theta_0;X_1)-G_1(0;X_1)\}\tilde
b_j|\mathcal{X}]\label{ross}
\\
\n &=&g(X_1)E\{\tilde b_j(a_1-\tilde a_j-\tilde
b_jX_{1j}^\t\theta_0)|\mathcal{X}\}+O[E\{(a_1-\tilde a_j-\tilde b_jX_{1j}^\t\theta_0)^2|\mathcal{X}\}].
\endy
Using the results on the asymptotic bias and variance of  $(\tilde a_j,\tilde b_j)$  in (\ref{welsh}),
we can see that
\beginn
E\{K_{1j}^\vartheta(a_1-\tilde a_j-\tilde b_jX_{1j}^\t\theta_0)^2\}=O(h\delta_\vartheta^2+n^{-1}),
\endn
Next we deal with the first term in (\ref{ross}). Using (\ref{lu1}),
\beginy
\n &&a_1-\tilde a_j-\tilde b_jX_{1j}^\t\theta_0=a_1-a_j+a_j-\tilde
a_j-\tilde
b_jX_{1j}^\t\theta_0\\
\n &&=\frac{1}{2}m''(X_j^\t\theta_0)\{(X_{1j}^\t\theta_0)^2\}
-\frac{1}{2}m''(X_j^\t\theta_0)h^2+O\{(X_{1j}^\t\theta_0)^3\}
\\\n&&-b_j\delta_\vartheta^\t\{(\nu/\mu)_\vartheta(X_j)-X_j\}
-b_j\delta_\vartheta^\t\{(\mu\nu'-\mu'\nu)/\mu^2\}_\vartheta(
X_j)X_{1j}^\t\theta_0\\
\n&&-h^2[\frac{1}{2}m^{''}(X_j^\t\theta_0)\{(f\mu)'/(fg)\}_\vartheta(X_j)
+\frac{1}{6}m^{(3)}(X_j^\t\theta_0)(f\mu)_\vartheta(X_j)]X_{1j}^\t\theta_0\\
\n&&+\{gf\}^{-1}_\vartheta(X_j)\frac{1}{nh}
\sum\limits_{i=1}^n\varphi_{ij}-\{gf\}^{-1}_\vartheta(X_j)
\{\frac{1}{nh^2}\sum\limits_{i=1}^n\tilde\varphi_{ij}
\}X_{1j}^\t\theta_0\\
&&+O\{(nh/\log n)^{-3/4}(1+\delta_\vartheta/h)+h^3\}\label{re}
\endy
where $\varphi_{ij},\ \tilde\varphi_{ij}$ are zero-mean $\rm IID$ random variables
\beginy
 E[K^\vartheta_{1j}X_{1j}\T_1]&=&E[K^\vartheta_{1j}g(X_1)X_{1j}\tilde b_1(a_1-\tilde
a_j-\tilde
b_jX_{1j}^\t\theta_0)]+o(h|\delta_\vartheta|+n^{-1/2}h)\label{xixi}\\
\n &=&-h\delta_\vartheta^\t
b^2_j\{(\nu/\mu)_\vartheta(X_j)-X_j\}\{\nu_\vartheta(X_j)-X_j\mu_\vartheta(X_j)\}
+o(h|\delta_\vartheta|+hn^{-1/2})
\endy
uniformly in $\vartheta$, where (\ref{welsh}) is used in the last
step.

As $P([\tilde a_j,\tilde b_j]$ $\ne [\hat a_j,\hat
b_j])=O\{(nh/\log^3 n)^{-1/4}\} $, we have similar to $\T_2$ in (\ref{win}),
\beginn
 E[K^\vartheta_{1j}X_{1j}\T_2]
=o(n^{-1/2}h)+o(h\delta_\vartheta),\quad E[K^\vartheta_{1j}X_{1j}\T_2]=o(n^{-1/2}h)+o(h\delta_\vartheta)
\endn
uniformly in $\vartheta.$
This together with (\ref{xixi}) yields (\ref{f1}).

To prove (\ref{f2}), first note that
\beginn
\varphi(Y_i-\hat a_j-\hat b_j\theta_0^\t {X}_{ij})-\varphi(\e_i)&=&[
\varphi(Y_i-\hat a_j-\hat b_j\theta_0^\t {X}_{ij})-\varphi(Y_i- a_j-
b_j\theta_0^\t {X}_{ij})] \\
&&+[ \varphi(Y_i-\ a_j- b_j\theta_0^\t {X}_{ij})-\varphi(\e_i)].
\endn
Let $\tilde Z_{ij}=K^\vartheta_{ij}  X_{ij}\{\varphi(Y_i-\ a_j-
b_j\theta_0^\t {X}_{ij})-\varphi(\e_i)\}$. By Lemma \ref{anhui}, it
suffices to show that
\beginy
&&\sum\limits_{i,j}b_j( \tilde Z_{ij}-E \tilde Z_{ij})=o(n^2h\delta_\vartheta)\label{niew}\\
&&\sum\limits_{j}\ (\hat
b_j-b_j)\sum_{i}\tilde Z_{ij}=o(n^2h\delta_\vartheta).
\label{shi}
\endy
Due to Borel-Cantelli Lemma, (\ref{niew}) can be further reduced to, for any $\epsilon>0$
\beginy
&&nP\{|\sum\limits_{i}b_j(\tilde  Z_{ij}-E\tilde  Z_{ij})|\ge \epsilon
nh\delta_\vartheta\}\mbox{ is summable over }n,\label{old}
\endy
which follows from the facts that  $\tilde Z_{ij}$ is
bounded, $E\tilde Z_{ij}^2=O(h^3+h\delta_\vartheta^2)$ and
Bernstein's inequality,
\beginn
P\{|\sum\limits_{i}( \tilde Z_{ij}-E\tilde  Z_{ij})|\ge \epsilon
nh\delta_\vartheta\}\le C\exp\Big\{-\frac{\epsilon^2
n^{2}h^2\delta_\vartheta^2}{nh^3+nh\delta_\vartheta^2+\epsilon
nh\delta_\vartheta}\Big\}=o(n^{-2}).
\endn
To prove (\ref{shi}),
we again use the  expansion of $\hat b_j-b_j$ given in (\ref{lu1}), i.e.
\beginn
\hat
b_j-b_j&=&h^2\Big[\frac{1}{2}m^{''}(X_j^\t\theta_0)\{(f\mu)'/(fg)\}_\vartheta(X_j)
+\frac{1}{6}m^{(3)}(X_j^\t\theta_0)\{(f\mu)/(fg)\}_\vartheta(X_j)\Big]\\
\n&&
+b_j\delta_\vartheta^\t\{(\mu\nu'-\mu'\nu)/\mu^2\}_\vartheta(X_j)+\frac{1}{nh^2}\sum\limits_{i=1}^n
\tilde\varphi_{ij}+O\{(nh/\log n)^{-3/4}/h\}
\endn
where $E\tilde\varphi_{ij}=0.$ If we denote by $C(X_j)$ the determinstic(bias) term in $\hat b_j-b_j$,
it is easy to see that  
 $\sum_{i,j}C(X_j)\tilde Z_{ij}=o(n^2h\delta_\vartheta)$ .
For
the stochastic part, write
\beginy
\sum
\limits_{j,i,l}\tilde Z_{ij}\tilde\varphi_{lj}
=\sum \limits_{i,j}\tilde Z_{ij}\tilde\varphi_{ij}
+\sum \limits_{j,i\ne l}\tilde Z_{ij}\tilde\varphi_{lj} \label{ce}
\endy
 We focus on the second term, as  the first term is relatively negligible. Let $c\equiv E\tilde Z_{ij}=O(h^3+h\delta_\vartheta^2)$, whence
the second term in (\ref{ce}) is $(nh^2)^{-1}\sum _{j}(\T_{1j}+c\T_{2j})$, where
\beginn
\T_{1j}=\sum \limits_{i< l}\{\tilde\varphi_{lj}
(\tilde Z_{ij}-c)+\tilde\varphi_{ij}(
\tilde Z_{lj}-c)\},\quad \T_{2j}=\sum \limits_{i<
l}(\tilde\varphi_{lj}+\tilde\varphi_{ij}).
\endn
By the second statement in Lemma 6.1 in Xia(2007), replacing $\theta$ there with $(\vartheta^\t,X_j^\t)^\t$, 
we know that with probabiltity $1,$ $\T_{1j}=O\{n\log n(h^3+h\delta_\vartheta^2)^{1/2}\}$
 uniformly in $\vartheta$ and $j$. On the
other hand, by law of the iterated logarithm for U-statistics in
Korolyuk et al (Lemma \ref{koro}), $\sum _{j}\T_{2j}=n^{3/2}(h\log\log n)^{1/2}\
a.s.$ Since $c=O(h^3+h\delta_\vartheta^2)$,
 we have 
 \beginn
 \frac{1}{nh^2}\sum _{j}(\T_{1j}+c\T_{2j})=\frac{1}{nh^2}O\{n^2\log n(h^3+h\delta_\vartheta^2)^{1/2}+n^{3/2}\log n)(h^3+h\delta_\vartheta^2)\}=o(n^2h\delta_\vartheta)
 \endn

 Proof of (\ref{garner})  can be done in exactly the same manner as
(\ref{shi}).\hspace{\fill}$\blacksquare$

The proof of (\ref{ricci1}) consists of the following two Lemmas.

\begin{Lemma}\label{wuhu}Let $R^*_{n2}(\theta)=\sum\limits_{i,j}K^\vartheta_{ij}\Big[\rho(Y_i-\hat
a_j-\hat b_j\theta^\t {X}_{ij})-\rho(Y_{ij})-\delta_\theta^\t
\varphi(Y_i-a_j-b_jX_{ij}^\t\theta_0)\hat b_jX_{ij}\Big]$. Then with
probability $1,$ we have
\beginy
(n^2ha^2_{n\vartheta})^{-1}[R_{n2}^*(\theta)-ER^*_{n2}(\theta)]=o(1).\label{ricci}
\endy
uniformly in $\vartheta.$
\end{Lemma}
\noindent{\bf Proof }  Define ${X}_{ix}={X}_i-{x},\
\mu_{ix}=(1,X_{ix}^\t)^\t,\ K_{ix}=K(X_{ix}^\t\vartheta/{h}),\
\beta({x})=[m(\theta_0^\t x),m'(\theta_0^\t x)\theta_0^\t]^\t$ and
$\varphi_{ni}({x};t)=\varphi(Y_{i};\mu_{ix}^\t\beta({x})+t)$. For
 any $\alpha,\ \beta\in \R^{d+1}$, let
\beginn
\Phi_{ni}({x};\alpha,\beta)&=&K_{ix}\Big[\rho\{Y_{i};\mu_{ix}^\t(\alpha+\beta+\beta({x}))\}
-\rho\{Y_{i};\mu_{ix}^\t(\beta+\beta({x}))\})-\varphi_{ni}({x};0)\mu_{ix}^\t\alpha\Big]\\
&=&K_{ix}\int\limits_{\mu_{ix}^\t\beta}^{\mu_{ix}^\t
(\alpha+\beta)}\{\varphi_{ni}({x};t)-\varphi_{ni}({x};0)\}dt
\endn
and
$R_{ni}({x};\alpha,\beta)=\Phi_{ni}({x};\alpha,\beta)-E\Phi_{ni}({x};\alpha,\beta).$
Apparently,
\beginn
K^\vartheta_{ij}\Big[\rho(Y_i-\hat a_j-\hat b_j\theta^\t
{X}_{ij})-\rho(Y_{ij})-\delta_\theta^\t
\varphi(Y_i-a_j-b_jX_{ij}^\t\theta_0)\hat b_jX_{ij}\Big]\equiv
\Phi_{ni}(X_j;\alpha,\beta)
\endn
with $\alpha=[0,\hat b_j\delta_\theta^\t ]^\t$ and $\beta=[\hat
a_j-a_j,(\hat b_j-b_j)\theta_0^\t ]^\t$. Let $[a_x,b_x]\equiv
[m(\theta_0^\t x),m'(\theta_0^\t x)] $ and ${\mathcal D} $ be any
compact subset of the support of $X$. For any $M>0$ and
$\vartheta\in \Theta_n,$ define
\beginn
&M^\vartheta_{n1}=Ca_{n\vartheta},\ \
 M^\vartheta_{n2}=C\{|\delta_\vartheta|+(nh/\log
n)^{-1/2}\}, & \\
& M^\vartheta_{n3}=C\{|\delta_\vartheta|+(nh/\log n)^{-1/2}/h\}, \ \
\  B_{n}^{(1)}=\{\alpha\in R^{d+1}|\alpha=[0,\alpha_1^\t]^\t,
|\alpha_1|\le M^\vartheta_{n1}\},&\\
 & B_{n}^{(2)}=\{\beta\in
R^{d+1}|\beta=[b_1,b_2\theta_0^\t ]^\t,|b_1|\le
M^\vartheta_{n2},|b_2|\le M^\vartheta_{n3}\}. &
\endn
 As $|\hat b_j\delta_\theta|\le Ca_{n\vartheta}$, $|\hat
a_j-a_j|=O\{|\delta_\vartheta|+(nh/\log n)^{-1/2}\}$ and $|(\hat
b_j-b_j)|=O\{|\delta_\vartheta|+(nh/\log n)^{-1/2}/h\}$,
 (\ref{ricci}) will follow if  for any $\epsilon>0$
\beginy
\sup\limits_{\scriptsize{x}\in {\mathcal D}}\sup\limits_{\scriptsize\begin{matrix}\alpha\in B_{n}^{(1)},\\
\beta\in B_{n}^{(2)}\end{matrix}}
|\sum\limits_{i=1}^{n}R_{ni}({x};\alpha,\beta)|\le \epsilon d_n\
a.s., \ \ \ d_n=nha_{n\vartheta}^2\label{RR}
\endy
This is done in a similar
 style as Lemma 4.2 in Kong et al(2008).  Cover
${\mathcal D}$ by a finite number $\T_n$ of cubes ${\mathcal
D}_k={\mathcal D}_{n,k}$ with side length $l_n=O\{h({nh/\log
n})^{-1/4}\}$ and centers $ {x}_k= {x}_{n,k}$. Write
\begin{align*}
\sup\limits_{ {x}\in {\mathcal D}}\sup\limits_{\scriptsize\begin{matrix}\alpha\in B_{n}^{(1)},\\
\beta\in B_{n}^{(2)}\end{matrix}} |\sum\limits_{i=1}^{n}R_{ni}(
{x};\alpha,\beta)|\le& \max\limits_{\small 1\le k\le \rm\T_n}
\sup\limits_{\scriptsize\begin{matrix}\alpha\in B_{n}^{(1)},\\
\beta\in B_{n}^{(2)}\end{matrix}}
\Big|\sum\limits_{i=1}^{n}R_{ni}(
{x}_k;\alpha,\beta)\Big|\\
&+\max\limits_{1\le k\le \T_n}\sup\limits_{ {x}\in {\mathcal D}_k}
\sup\limits_{\scriptsize\begin{matrix}\alpha\in B_{n}^{(1)},\\
\beta\in B_{n}^{(2)}\end{matrix}}
\Big|\sum\limits_{i=1}^{n}\Big\{\Phi_{ni}( {x}_k;\alpha,\beta)-\Phi_{ni}( {x};\alpha,\beta)\Big\}\Big|\\
&+\max\limits_{1\le k\le \T_n}\sup\limits_{ {x}\in {\mathcal D}_k}
\sup\limits_{\scriptsize\begin{matrix}\alpha\in B_{n}^{(1)},\\
\beta\in B_{n}^{(2)}\end{matrix}}
\Big|\sum\limits_{i=1}^{n}\Big\{E\Phi_{ni}( {x}_k;\alpha,\beta)-E\Phi_{ni}( {x};\alpha,\beta)\Big\}\Big|\\
\equiv&Q_1+Q_2+Q_3.
\end{align*}
In Lemma \ref{maa}, we will prove that $Q_2=o(d_n),\  a.s.$, whence
 $ Q_3\le EQ_2=o(d_n)$. It remains to show that
$Q_1\le \epsilon d_n/3\ a.s.$, which can be done following a similar
proof style as in Lemma 4.2 in Kong et al (2008).

Partition $B_n^{(i)},\ i=1,2$ into a sequence of sub rectangles
$D_{1}^{(i)},\cdots,D_{J_1}^{(i)},\ i=1,2,$ such that for all $ 1\le
j_1\le J_1\le M^{d+1}\ (M=\epsilon ^{-1})$ and for all$
\alpha,\alpha' \in D_{j_1}^{(1)}$, we have $ |\alpha-\alpha'|\le
M_{n1}^{\vartheta}/M; $ for  all $ \beta=[b_1,b_2\theta_0^\t ]^\t
,\beta'=[b_1',b_2'\theta_0^\t ]^\t\in D_{j_1}^{(2)} $, we have $
|b_1-b_1'|\le M^\vartheta_{n2}/M,|b_2-b_2'|\le M^\vartheta_{n3}/M $.
 Choose a point $\alpha_{j_1}\in D_{j_1}^{(1)}$ and $b_{k_1}\in
D_{k_1}^{(2)}$, $ 1\le j_1,k_1\le J_1$. Then for any $x,$
\beginy
\n \sup_{\scriptsize\begin{matrix}\alpha\in B_n^{(1)}\\
\beta\in B_n^{(2)}\end{matrix}}
|\sum\limits_{i}R_{ni}(
{x};\alpha,\beta)|&\le &\max\limits_{\scriptsize
1\le j_1, k_1\le J_1}
\sup\limits_{\scriptsize\begin{matrix}\alpha\in D_{j_1}^{(1)},\\
\beta\in D_{k_1}^{(2)}\end{matrix}}|\sum\limits_{i=1}^{n}
\{R_{ni}(x;\alpha_{j_1},b_{k_1})-R_{ni}(
{x};\alpha,\beta)\}|\\
&&+\max\limits_{\scriptsize
 1\le j_1,k_1\le J_1}
|\sum\limits_{i=1}^{n}R_{ni}(x;\alpha_{j_1},\beta_{k_1})|=H_{n1}+H_{n2}.\label{check}
\endy
 We first show that any $\epsilon>0$
\beginy
\T_n P\Big\{H_{n2}\ge \frac{\epsilon d_n}{2}\Big\}\le \T_n
J_1^2P\Big\{|\sum\limits_{i=1}^{n}R_{ni}(x;\alpha_{j_1},\beta_{k_1})|\ge
\frac{\epsilon d_n}{3}\Big\}=O(n^{-a}),\label{nn}
\endy
for some $a>1.$ By Bernstein's Inequality and the fact that
$|R_{ni}(x;\alpha_{j_1},\beta_{k_1})|\le Ca_{n\vartheta}$  and $\Var
\{R_{ni}(x;\alpha_{j_1},\beta_{k_1})\}=O[nha_{n\vartheta}^2\{a_{n\vartheta}+(nh/\log
n)^{-1/2}\}]$, we have
\beginn
T_nJ_1^2P\Big\{|\sum\limits_{i=1}^{n}R_{ni}(x;\alpha_{j_1},\beta_{k_1})|\ge
\frac{\epsilon d_n}{3}\Big\}=T_nJ_1^2\exp[-\epsilon^2
nha_{n\vartheta}\{1+a_{n\vartheta}(nh/\log n)^{1/2})\}]=O(n^{-a}),
\endn
for some $a>1.$ Therefore, (\ref{nn}) holds.

We next consider $H_{n1}$. For each $j_1=1,\cdots,J_1$ and $i=1,2$,
partition each rectangle $D_{j_1}^{(i)}$ further into a sequence of
subrectangles $D_{j_1,1}^{(i)},\cdots,D_{j_1,J_2}^{(i)}$. Repeat
this process recursively as follows. Suppose after the $l$th round,
we get a sequence of rectangles $D_{j_1,j_2,\cdots,j_l}^{(i)}$ with
$1\le j_k\le J_k,\ 1\le k\le l$, then in the $(l+1)$th round, each
rectangle $D_{j_1,j_2,\cdots,j_l}^{(i)}$ is partitioned into a
sequence of subrectangles
$\{D_{j_1,j_2,\cdots,j_l,j_{l+1}}^{(i)},1\le j_{l}\le J_{l}\}$ such
that for all $1\le j_{l+1}\le J_{l+1}$ and for all $  a,a'\in
D_{j_1,j_2,\cdots,j_l,j_{l+1}}^{(i)}, $ we have $ |a-a'|\le
M^\vartheta_{n1}/M ^{l+1}$; and for all $\beta=[b_1,b_2\theta_0^\t
]^\t ,\beta'=[b_1',b_2'\theta_0^\t ]^\t\in
D_{j_1,j_2,\cdots,j_l,j_{l+1}}^{(2)}, |b_1-b_1'|\le
{M^\vartheta_{n2}}/{M^{l+1}},|b_2-b_2'|\le
{M^\vartheta_{n3}}/{M^{l+1}}, $ where $J_{l+1}\le M^{d+1}$. Repeat
 this process after the $(\L_n+2)$th round, with $\L_n$ being the
largest integer such that
\beginy
n(2/M)^{L_n}>d_n/M^\vartheta_{n2}.\ \label{utrecht1}
\endy
 Let $D_{l}^{(i)},\ i=1,2$, denote the set of all
subrectangles of $D_{0}^{(i)}$ after the $l$th round of partition
and a typical element $D_{j_1,j_2,\cdots,j_l}^{(i)}$ of
$D_{l}^{(i)}$ is denoted as $D_{(j_l)}^{(i)}$. Choose a point
$\alpha_{(j_l)}\in D_{(j_l)}^{(1)}$ and $\beta_{(j_l)}\in
D_{(j_l)}^{(2)}$. Define
\beginn
&&\hspace{-.3cm}V_l=\sum\limits_{\tiny\begin{matrix}(j_{l+1})\\
(k_{l+1})\end{matrix}}P\Big\{\Big|\sum\limits_{i=1}^{n}
\{R_{ni}(x;\alpha_{(j_l)},\beta_{(k_l)})-R_{ni}(x;\alpha_{(j_{l+1})},\beta_{(k_{l+1})})\}\Big|
\ge \frac{\e d_n}{2^{l+1}}\Big\},\ 1\le l\le \L_n+1,\\
&&\hspace{-.3cm}Q_l=\sum\limits_{\tiny\begin{matrix}(j_{l})\\
(k_{l})\end{matrix}}
P\Big\{\sup\limits_{\tiny\begin{matrix}\alpha\in D_{(j_l)}^{(1)},\\
\beta\in D_{(k_l)}^{(2)}\end{matrix}}\Big|\sum\limits_{i=1}^{n}
\{R_{ni}(x;\alpha_{(j_l)},\beta_{(k_l)})-R_{ni}(x;\alpha,\beta)\}\Big|\ge
\frac{\e d_n}{2^l}\Big\}, \ 1\le l\le \L_n+2.
\endn
Then $Q_l\le V_l+Q_{l+1},\ 1\le l\le L_n+1.$ On the other hand, it
is easy to see that for any $\alpha \in D_{(j_{L_n+2})}^{(1)}$ and $
\beta\in D_{(k_{L_n+2})}^{(2)}$,
\beginn
n|R_{ni}(x;\alpha_{(j_{L_n+2})},\beta_{(k_{L_n+2})})-R_{ni}(x;\alpha,\beta)|\le
n{M_{n2}^\vartheta}/{M^{L_n+2}}\le \epsilon d_n/2^{L_n+2}
\endn
due to the choice of $L_n$ specified in (\ref{utrecht1}). Therefore,
$Q_{L_n+2}=0$ and it remains to show that
\beginy
T_nP\{H_{n1}\ge  \frac{\epsilon d_n}{2}\}\le T_nJ_1^2 Q_1\le
T_nJ_1^2 \sum_{l=1}^{L_n+1} V_l=O(n^{-a}), \mbox{ for some }a>1.
\label{smoke}
\endy
To find upper bound for $V_l,\ 1\le l\le L_n+1$, we again apply
Bernstein's inequality.  As
\beginn
&&|R_{ni}(x;\alpha_{(j_l)},\beta_{(k_l)})-R_{ni}(x;\alpha_{(j_{l+1})},\beta_{(k_{l+1})})|\\
&\le&
C\{|\alpha_{(j_l)}-\alpha_{(j_{l+1})}|+
|\beta_{(k_l)}-\beta_{(k_{l+1})}|(\delta_\vartheta+h)\}\equiv
{M_{n2}^\vartheta}/{M^{l}},
\endn
\vspace{-1cm}
\beginn
&E|R_{ni}(x;\alpha_{(j_l)},\beta_{(k_l)})-R_{ni}(x;\alpha_{(j_{l+1})},\beta_{(k_{l+1})})|^2\le
h (M_{n2}^\vartheta)^3/{M^{l}},&
\endn
 we have
\beginn
V_l\le \Big(\prod\limits_{j=1}^{l+1}J_j^2\Big)
\exp[-\e^2nh\{1+a_{n\vartheta}(nh/\log n)^{1/2}\}],\
\endn
 and (\ref{smoke}) thus holds.
 This together with (\ref{nn}) completes the proof.\hspace{\fill}$\blacksquare$

\begin{Lemma}\label{anhui} Let $Z_{ij}=K_{ij}[ \varphi(Y_i- a_j- b_j\theta_0^\t {X}_{ij})-
\varphi(Y_i-\hat a_j-\hat b_j\theta_0^\t {X}_{ij})]\hat
b_jX_{ij} $. Then
\beginy
\sum_{i,j}Z_{ij}-EZ_{ij}=o(n^2ha_{n\vartheta}).
\label{kosher}
\endy
\end{Lemma}

\noindent{\bf Proof}
 As $\hat a_j-a_j=O(a_{n\vartheta})$, $(\hat
b_j-b_j)=O\{a_{n\vartheta}+(nh/\log n)^{1/2}/h\}$ and for any
$\epsilon>0$,
\beginn
P\Big\{|\sum_{i,j}Z_{ij}-EZ_{ij}|\ge \epsilon
n^2ha_{n\vartheta}\Big\}\le nP\Big\{|\sum_{i}Z_{ij}-EZ_{ij}|\ge
\epsilon nha_{n\vartheta}\Big\}
\endn
 then (\ref{kosher}) would follow if we could show that for any $x$,
 \beginy
P\Big\{ \sup_{\scriptsize\begin{matrix}\alpha\in B_n^{(1)}\\
\beta\in B_n^{(2)}\end{matrix}} |\sum\limits_{i}R_{ix}(a,b)|\ge
\epsilon nha_{n\vartheta}\Big\}=O(n^{-a})\ \mbox{for some
}a>2,\label{maas}
\endy
where $B_n^{(1)}=\{a\in R: |a-a_x|\le ca_{n\vartheta}\},\
B_n^{(2)}=\{b\in R: |b-b_x|\le c\{a_{n\vartheta}+(nh/\log
n)^{1/2}/h\}\},\ a_x=m(\theta_0^\t x),\ b_x=m'(\theta_0^\t x)$,
$R_{ix}(a,b)=Z_{ix}(a,b)-EZ_{ix}(a,b)$, $ K_{ix}=K(X_{ix}^\t\vartheta/h) $ and $
Z_{ix}(a,b)=K_{ix} X_{ix}[ \varphi(Y_i- a_x- b_x\theta_0^\t
{X}_{ix})- \varphi(Y_i- a- b \theta_0^\t {X}_{ix})]$.
To this end, partition $B_n^{(i)},\ i=1,2$ into a sequence of sub
rectangles $D_{1}^{(i)},\cdots,D_{J_1}^{(i)},\ i=1,2$ such that
\beginn
|D_{j_1}^{(i)}|=\sup\Big\{|a-a'|:a,a' \in D_{j_1}^{(i)}\Big\}\le
M_{n}^{(i)}/M, \ \ 1\le j_1\le J_1,
\endn
where $M_{n}^{(1)}=ca_{n\vartheta}$,
$M_{n}^{(2)}=c\{a_{n\vartheta}+(nh/\log n)^{1/2}/h\}, M\equiv \epsilon^{-1} $ and $J_1\le
M$. Choose a point $a_{j_1}\in D_{j_1}^{(1)}$ and $b_{k_1}\in
D_{k_1}^{(2)}$. Then
\beginy
\n \sup_{\scriptsize\begin{matrix}a\in B_n^{(1)}\\
b\in B_n^{(2)}\end{matrix}} |\sum\limits_{i}R_{ix}(a,b)|&\le
&\max\limits_{\scriptsize 1\le j_1, k_1\le J_1}
\sup\limits_{\scriptsize\begin{matrix}a\in D_{j_1}^{(1)},\\
b\in D_{k_1}^{(2)}\end{matrix}}|\sum\limits_{i=1}^{n}
\{R_{ix}(a_{j_1},b_{k_1})-R_{ix}(a,b)\}|\\
&&+\max\limits_{\scriptsize
 1\le j_1,k_1\le J_1}
|\sum\limits_{i=1}^{n}R_{ix}(a_{j_1},b_{k_1})|\equiv H_{n1}+H_{n2}.\label{check}
\endy
We first consider $H_{n2}$.
\beginn
P\Big\{H_{n2}\ge \frac{\e nha_{n\vartheta}}{2}\Big\}\le
J_1^2P\Big\{|\sum\limits_{i=1}^{n}R_{ix}(a_{j_1},b_{k_1})|\ge
\frac{\epsilon nha_{n\vartheta}}{2}\Big\}
\endn
As $R_{ix}(a_{j_1},b_{k_1})$ is bounded and $\Var
\{R_{ix}(a_{j_1},b_{k_1})\}=O\{h(a_{n\vartheta}+(nh/\log
n)^{-1/2}\}$, then by Bernstein's inequality we have
\beginn
J_1^2P\Big\{|\sum\limits_{i=1}^{n}R_{ix}(a_{j_1},b_{k_1})|\ge
\frac{\epsilon nha_{n\vartheta}}{2}\Big\}\le CJ_1^2\exp\{-\epsilon^2
n^{1/2}h^{3/2}\}=O(n^{-a}),
\endn
for some $a>2.$

\noindent We next consider $H_{n1}$. For each $j_1=1,\cdots,J_1$ and
$i=1,2$, partition each rectangle $D_{j_1}^{(i)}$ further into a
sequence of subrectangles
$D_{j_1,1}^{(i)},\cdots,D_{j_1,J_2}^{(i)}$. Repeat this process
recursively as follows. Suppose after the $l$th round, we get a
sequence of rectangles $D_{j_1,j_2,\cdots,j_l}^{(i)}$ with $1\le
j_k\le J_k,\ 1\le k\le l$, then in the $(l+1)$th round, each
rectangle $D_{j_1,j_2,\cdots,j_l}^{(i)}$ is partitioned into a
sequence of subrectangles
$\{D_{j_1,j_2,\cdots,j_l,j_{l+1}}^{(i)},1\le j_{l}\le J_{l}\}$ such
that
\beginn
|D_{j_1,j_2,\cdots,j_l,j_{l+1}}^{(i)}|=\sup\Big\{|a-a'|:a,a'\in
D_{j_1,j_2,\cdots,j_l,j_{l+1}}^{(i)}\Big\}\le M_{n}^{(i)}/M ^{l+1},\
1\le j_{l+1}\le J_{l+1},
\endn
where $J_{l+1}\le M$. End this process after the $(\L_n+2)$th round,
with $\L_n$ being the smallest integer such that
\beginy
(2/M)^{L_n}>a_{n\vartheta}/M_{n\vartheta}^{(2)}\ [\mbox{which means
}2^{L_n}\le \{M_{n\vartheta}^{(2)}/a_{n\vartheta}\}^{\log
{(M/2)}/\log 2} ]. \label{utrecht}
\endy
 Let $D_{l}^{(i)},\ i=1,2$, denote the set of all
subrectangles of $D_{0}^{(i)}$ after the $l$th round of partition
and a typical element $D_{j_1,j_2,\cdots,j_l}^{(i)}$ of
$D_{l}^{(i)}$ is denoted as $D_{(j_l)}^{(i)}$. Choose a point
$a_{(j_l)}\in D_{(j_l)}^{(1)}$ and $b_{(j_l)}\in D_{(j_l)}^{(2)}$
and define
\beginn
&&\hspace{-.3cm}V_l=\sum\limits_{\tiny\begin{matrix}(j_{l})\\
(k_{l})\end{matrix}}P\Big\{\Big|\sum\limits_{i=1}^{n}
\{R_{ix}(a_{j_l},b_{k_l})-R_{ix}(a_{j_{l+1}},b_{k_{l+1}})\}\Big|
\ge \frac{\epsilon nha_{n\vartheta}}{2^{l+1}}\Big\},\ 1\le l\le \L_n+1,\\
&&\hspace{-.3cm}Q_l=\sum\limits_{\tiny\begin{matrix}(j_{l})\\
(k_{l})\end{matrix}}
P\Big\{\sup\limits_{\tiny\begin{matrix}a\in D_{(j_l)}^{(1)},\\
b\in D_{(k_l)}^{(2)}\end{matrix}}\Big|\sum\limits_{i=1}^{n}
\{R_{ix}(a_{j_l},b_{k_l})-R_{ix}(a,b)\}\Big|\ge \frac{\epsilon
nha_{n\vartheta}}{2^l}\Big\}, \ 1\le l\le \L_n+2.
\endn
Then $Q_l\le V_l+Q_{l+1},\ 1\le l\le L_n+1.$ We first give a bound
for $V_l,\ 1\le l\le L_n+1$. As
$R_{ix}(a_{j_l},b_{k_l})-R_{ix}(a_{j_{l+1}},b_{k_{l+1}})$ is bounded
and
\beginn
E|R_{ix}(a_{j_l},b_{k_l})-R_{ix}(a_{j_{l+1}},b_{k_{l+1}})|^2\le
h\{a_{n\vartheta}+(nh/\log n)^{-1/2}\}/M^{l+1},
\endn
applying Bernstein's inequality and using (\ref{utrecht}), we have
\beginy
V_l\le \Big(\prod\limits_{j=1}^{l+1}J_j^2\Big)
\exp[-\epsilon^2nh\min\{a_{n\vartheta},a^2_{n\vartheta}(nh/\log
n)^{1/2}\}]\le
\Big(\prod\limits_{j=1}^{l+1}J_j^2\Big)\exp(-\epsilon^2n^{1/2}h^{3/2}).
\label{ff}
\endy
We now focus on $Q_{L_n+2}$. Recall the definition of $Z_{ix}(a,b)$
\beginn
Z_{ix}(a,b)=K_{ix}[ \varphi(Y_i- a_x- b_x\theta_0^\t {X}_{ix})-
\varphi(Y_i- a- b \theta_0^\t {X}_{ix})] X_{ix}.
\endn
For any $a\in D_{(j_l)}^{(1)}$ and $ b\in D_{(k_l)}^{(2)}$, let
$I_i^{a,b}=1$, if there is a discontinuity point of $\varphi(.)$
between $Y_i-a_{j_l}-b_{k_l} \theta_0^\t {X}_{ix}$ and $Y_i- a- b
\theta_0^\t {X}_{ix}$ and $I_i^{a,b}=0$ otherwise. Write
\beginn
R_{ix}(a_{j_l},b_{k_l})-R_{ix}(a,b)=
\{R_{ix}(a_{j_l},b_{k_l})-R_{ix}(a,b)\}I_i^{a,b}+
\{R_{ix}(a_{j_l},b_{k_l})-R_{ix}(a,b)\}(1-I_i^{a,b}).
\endn
Then we have
$|\{R_{ix}(a_{j_l},b_{k_l})-R_{ix}(a,b)\}(1-I_i^{a,b})|\le
C\{a_{n\vartheta}+(nh/\log n)^{-1/2}\}/M^{l}$ and specifically for
$l=\L_n+2$
\beginn
&&P\Big\{\sup\limits_{\tiny\begin{matrix}a\in D_{(j_l)}^{(1)},\\
b\in D_{(k_l)}^{(2)}\end{matrix}}\Big|\sum\limits_{i=1}^{n}
\{R_{ix}(a_{j_l},b_{k_l})-R_{ix}(a,b)\}(1-I_i^{a,b})\Big|\ge
\frac{\epsilon nha_{n\vartheta}}{2^{\L_n+3}}\Big\}\\
&&\hspace{3cm}\le P\Big\{\sum\limits_{i=1}^{n} U_i\ge
\frac{1}{8}{Mnh}\Big\}\le P\Big\{\sum\limits_{i=1}^{n} U_i-EU_i\ge
\frac{Mnh}{16}\Big\}
\endn
where $U_i=I\{|X_{ix}^\t\vartheta|\le h\}$ and the first inequality
is due to (\ref{utrecht}). By Bernstein's inequality, this in turn
implies that for $l=L_n+2$
\beginy
\Big(\prod\limits_{j=1}^{l+1}J_j^2\Big)
P\Big\{\sup\limits_{\tiny\begin{matrix}a\in D_{(j_l)}^{(1)},\\
b\in D_{(k_l)}^{(2)}\end{matrix}}\Big|\sum\limits_{i=1}^{n}
\{R_{ix}(a_{j_l},b_{k_l})-R_{ix}(a,b)\}(1-I_i^{a,b})\Big|\ge
\frac{\epsilon
nha_{n\vartheta}}{2^{\L_n+3}}\Big\}=O(n^{-a}),\label{h}
\endy
for some $a>2.$ Now we have to show similar result for
\beginn
\Big(\prod\limits_{j=1}^{l+1}J_j^2\Big)
P\Big\{\sup\limits_{\tiny\begin{matrix}a\in D_{(j_l)}^{(1)},\\
b\in D_{(k_l)}^{(2)}\end{matrix}}\Big|\sum\limits_{i=1}^{n}
\{R_{ix}(a_{j_l},b_{k_l})-R_{ix}(a,b)\}I_i^{a,b}\Big|\ge
\frac{\epsilon nha_{n\vartheta}}{2^{\L_n+3}}\Big\},\ l=L_n+2.
\endn
Note that for any $a\in D_{(j_l)}^{(1)}$ and $ b\in
D_{(k_l)}^{(2)}$, $I_i^{a,b}\le I\{Y_i\in S_i\}$, where
\beginn
S_i=[a_{j_l}+b_{k_l}\theta_0^\t X_{ix}-CM_{n}^{(2)}/M^l,
a_{j_l}+b_{k_l}\theta_0^\t X_{ix}+CM_{n}^{(2)}/M^l],
\endn
which is independent of $a,b$. Let $U_i=I\{|X_{ix}^\t\vartheta|\le
h\}I\{Y_i\in S_i\}$. As $R_{ix}(a_{j_l},b_{k_l})-R_{ix}(a,b)$ is
bounded, we have for $l=L_n+2,$
\beginy
\n&&P\Big\{\sup\limits_{\tiny\begin{matrix}a\in D_{(j_l)}^{(1)},\\
b\in D_{(k_l)}^{(2)}\end{matrix}}\Big|\sum\limits_{i=1}^{n}
\{R_{ix}(a_{j_l},b_{k_l})-R_{ix}(a,b)\}I_i^{a,b}\Big|\ge
\frac{\epsilon
nha_{n\vartheta}}{2^{\L_n+3}}\Big\}\\
&&\le P\Big\{\sum\limits_{i=1}^{n} U_i\ge \frac{\epsilon
nha_{n\vartheta}}{C2^{\L_n+2}}\Big\}\le P\Big\{\sum\limits_{i=1}^{n}
U_i-EU_i\ge \frac{\epsilon
nha_{n\vartheta}}{C2^{\L_n+4}}\Big\},\label{Bern}
\endy
where the second inequality is due to (\ref{utrecht}). Applying
Bernstein's inequality to the right hand side of (\ref{Bern}) and by
(\ref{utrecht}), we have
\beginn
\Big(\prod\limits_{j=1}^{l+1}J_j^2\Big)
P\Big\{\sup\limits_{\tiny\begin{matrix}a\in D_{(j_l)}^{(1)},\\
b\in D_{(k_l)}^{(2)}\end{matrix}}\Big|\sum\limits_{i=1}^{n}
\{R_{ix}(a_{j_l},b_{k_l})-R_{ix}(a,b)\}I_i^{a,b}\Big|\ge
\frac{\epsilon nha_{n\vartheta}}{2^{\L_n+3}}\Big\}=O(n^{-a}),\mbox{
for }l=L_n+2
\endn
for some $a>2.$ This together with (\ref{h}) implies that
$Q_{L_n+2}=O(n^{-a})$ for some $a>2.$ Therefore, based on
(\ref{ff}), we have
\beginn
P\Big\{H_{n2}\ge \frac{\epsilon nha_{n\vartheta}}{2}\Big\}\le Q_1\le
\sum\limits_{l=1}^{L_n+1}V_l+Q_{L_n+2}=O(n^{-a}),
\endn
for some $a>2.$ $\hspace{\fill}\blacksquare$

\begin{Lemma} \label{maa} For all large enough
$M>0$, $Q_2\le Md_n\ a.s.$, where
 \beginn d_n=nha_{n\vartheta}^2
l_n/h\{1+a_{n\vartheta}^{-1}(nh/\log
n)^{-1/2}\}=o(nha^2_{n\vartheta}),
\endn

\end{Lemma}
\noindent{\bf Proof} Let ${X}_{ik}={X}_i-x_k,\
\mu_{ik}=(1,X_{ik}^\t)^\t,\ K_{ik}=K(X_{ik}^\t\vartheta/{h})$ and
write
$\Phi_{ni}({x}_k;\alpha,\beta)-\Phi_{ni}(x;\alpha,\beta)=\xi_{i1}+\xi_{i2}+\xi_{i3}$,
where
\beginn
&\xi_{i1}=\Big(K_{ik}\mu_{ik}-
K_{ix}\mu_{ix}\Big)^\t\alpha\int_0^1\left\{\varphi_{ni}(x_k;\mu_{ik}^\t(\beta+\alpha
t))
-\varphi_{ni}(x_k;0)\right\}dt,&\\
&\xi_{i2}=K_{ix}\mu_{ix}^\t\alpha\int_0^1\left\{\varphi_{ni}(x_k;\mu_{ik}^\t
(\beta+\alpha t))-\varphi_{ni}(x;\mu_{ix}^\t(\beta+\alpha t))\right\}dt,&\\
& \xi_{i3}=K_{ix}\mu_{ix}^\t\alpha\{\varphi_{ni}(x;0)-\varphi_{ni}(x_k;0)\}. &
\endn
Then $P(Q_2> M^{3/2}d_n/3)\le \T_n(P_{n1}+P_{n2}+P_{n3})$, where
\beginn
P_{nj}\equiv \max\limits_{1\le k\le \T_n}P\Big(\sup\limits_{ {x}\in
{\mathcal D}_k}
\sup\limits_{\scriptsize\begin{matrix}\alpha\in B_n^{(1)},\\
\beta\in B_n^{(2)}\end{matrix}} |\sum\limits_{i=1}^{n}\xi_{ij}|\ge
{M^{3/2}d_n}/{9}\Big),\ j=1,2,3.
\endn
Based on Borel-Cantelli lemma, $Q_2\le M^{3/2}d_n $ almost surely,
if  $\sum_{n}\T_nP_{nj}<\infty,\ j=1,2,3$. Again this can be
accomplished through similar approach in Lemma 5.1 in Kong et
al(2008). We only deal with $P_{nj}$ to illustrate.

First note that  if $\xi_{i1}\ne 0$,then either $K_{ik}\ne 0$ or
$K_{ix}\ne 0$. Without loss of generality, suppose $K_{ik}\ne 0,$
i.e. $ |X_{ix}^\t\vartheta|\le h$, whence $|X_{ix}^\t\theta_0|\le
h+|\delta_\vartheta|$ and $|\mu_{ik}^\t(\beta+\alpha t)|\le
C\{M_{n\vartheta}^{(1)}+M_{n\vartheta}^{(2)}\}$.

For any fixed $ \alpha\in B_n^{(1)}$ and $ \beta\in B_n^{(2)}$, let
$I^{\alpha,\beta}_{ik}=1$. If there exists some  $t\in [0,1]$, such
that there are discontinuity points of $\varphi(Y_{i}-a)$ between
$\mu_{ik}^\t(\beta( {x}_k)+\beta+\alpha t))$ and
$\mu_{ik}^\t\beta_p( {x}_k)$; and $I^{\alpha,\beta}_{ik}=0$,
otherwise. Write
$\xi_{i1}=\xi_{i1}I^{\alpha,\beta}_{ik}+\xi_{i1}(1-I^{\alpha,\beta}_{ik})$.
As  $|(K_{ik}\mu_{ik}- K_{ix}\mu_{ix})^\t\alpha|\le
CM_{n\vartheta}^{(1)}l_n/h$ and $|\mu_{ik}^\t(\beta+\alpha t)|\le
CM^{(2)}_{n\vartheta}$, we have
\beginn
|\xi_{i1}(1-I^{\alpha,\beta}_{ik})|\le
CM^{1}_{n\vartheta}M^{2}_{n\vartheta} l_n/h=o(a_{n\vartheta}^2)
\endn
uniformly in $i,\alpha$, $\beta$ and $ {x}\in {\mathcal D}_k$, if
$nh^3/\log n^3\to \infty$. Let $U_{ik}=I\{| {X}^\t_{ik}\vartheta|\le
2h\}$. As $\xi_{i1}=\xi_{i1}U_{ik}$ (because $l_n=o(h)$), we have
\beginy
\n P\Big(\sup\limits_{\scriptsize\begin{matrix}\alpha\in B_n^{(1)},\\
\beta\in B_n^{(2)}\end{matrix}}\sup\limits_{ {x}\in {\mathcal
D}_k}\Big|\sum\limits_{i=1}^{n}\xi_{i1}(1-I^{\alpha,\beta}_{ik})\Big|>\frac{Md_n}{18}\Big)&\le&
P\Big( \sum\limits_{i=1}^{n}U_{ik}>\frac{Mnh}{18C}\Big)\\
&\le& P\Big( |\sum\limits_{i=1}^{n}U_{ik}-EU_{ik}|>\frac{M
nh}{36C}\Big), \label{mul}
\endy
where the second inequality follows from the fact that
$EU_{ik}=O(h)$. We can then apply to (\ref{mul}) Bernstein's
inequality for independent data or Lemma 5.4 in Kong et al (2008) for
dependent case, to obtain the below result
\beginy
\T_nP\Big(\sup\limits_{\scriptsize\begin{matrix}\alpha\in B_n^{(1)},\\
\beta\in
B_n^{(2)}\end{matrix}}\Big|\sum\limits_{i=1}^{n}\xi_{i1}(1-I^{\alpha,\beta}_{ik})\Big|>Md_n/18\Big)
\mbox{ is summable over }n,\label{zz}
\endy
whence  $\sum_{n}\T_nP_{n1}<\infty,$ is equivalent to
\beginy
\T_nP\Big(\sup\limits_{\scriptsize\begin{matrix}\alpha\in B_n^{(1)},\\
\beta\in
B_n^{(2)}\end{matrix}}\Big|\sum\limits_{i=1}^{n}\xi_{i1}I^{\alpha,\beta}_{ik}\Big|>Md_n/18\Big)
\mbox{ is summable over }n.\label{ggg}
\endy
To this end, first note that  $I^{\alpha,\beta}_{ik}\le I\{\e_i\in
S^{\alpha,\beta}_{i;k}\}$, where
 \beginn
S^{\alpha,\beta}_{i;k}&=&\bigcup\limits_{j=1}^m\bigcup\limits_{t\in
[0,1]}[a_j-A( {X}_i, {x}_k)
+\mu_{ik}^\t(\beta+\alpha t),a_j-A( {X}_i, {x}_k)]\\
&\subseteq& \bigcup\limits_{j=1}^m [a_j-CM_{n\vartheta}^{(2)},
a_j+CM_{n\vartheta}^{(2)}]\equiv D_n,
\ \mbox{ for some }C>0,\label{shen}\\
A( {x}_1,
{x}_2)&=&m(x_1^\t\theta_0)-m(x_2^\t\theta_0)-m'(x_1^\t\theta_0)(x_1-x_2)^\t\theta_0,
\endn
where in the derivation of $S^{\alpha,\beta}_{i;k}\subseteq D_n$, we
have used the fact that $| {X}_{ik}|\le 2h,\
\mu_{ik}^\t(\beta+\alpha t)=O(M_n^{(2)})$ and $A( {X}_i,
{x}_k)=O(h^2+|\delta_\vartheta|^2)=o(M_n^{(2)})$ uniformly in $i$.
As $I^{\alpha,\beta}_{ik}\le I\{\e_i\in D_n\}$, we have
$|\xi_{i1}|I^{\alpha,\beta}_{ik}\le |\xi_{i1}| U_{ni}, $ where
$U_{ni}\equiv  I(| {X}_{ik}|\le 2h)I\{\e_i\in D_n\}$, which is
independent of the choice of $\alpha$ and $\beta.$
 Therefore,
\begin{align}
\n P\Big(\sup\limits_{\scriptsize\begin{matrix}\alpha\in B_n^{(1)},\\
\beta\in
B_n^{(2)}\end{matrix}}\Big|\sum\limits_{i=1}^{n}\xi_{i1}I^{\alpha,\beta}_{ik}\Big|>Md_n/18\Big)&\le
 P\Big(\sum\limits_{i=1}^{n}
U_{ni}>MnhM_n^{(2)}/(18C)\Big)\\
&\le
P\Big(\sum\limits_{i=1}^{n}(U_{ni}-EU_{ni})>\frac{MnhM_n^{(2)}}{36C}\Big),\label{ff}
\end{align}
where  the first inequality is because $|\xi_{i1}|\le
CMa_{n\vartheta}l_n/h$ and the second one  because $EU_{ni}=O(
hM_n^{(4)})$.
  Similar to (\ref{mul}), we could apply either Bernstein's
inequality for independent data or in dependent case Lemma 5.4 in
Kong et al (2008) to see that
 (\ref{ggg}) indeed holds.\hspace{\fill}$\blacksquare$

%Recall that
%\beginn
%S_2&=&\int \{m'({X}^\t\theta_0)\}^2
%\omega_{\theta_0}({X})f_{\theta_0}(X)dX,\\
%\omega_{\vartheta}({x})&=&E\{g(X)(X-x)(X-x)^\t|X^\t\vartheta=x^\t\vartheta\}.\\
%\Omega_0&\equiv& \lim_{n\to \infty}{n}^{-1}
%\sum_{j}b^2_j\mu_{\theta_0}(X_j)\{(\nu/\mu)_{\theta_0}(X_j)-X_j\}\{(\nu/\mu)_{\theta_0}(X_j)-X_j\}^\t\\
%&=&\int
%\{m'({X}^\t\theta_0)\}^2\mu_{\theta_0}(X)\{(\nu/\mu)_{\theta_0}(X)-X\}\{(\nu/\mu)_{\theta_0}(X)-X\}^\t
%d{F_{\theta_0}(X)}
%\endn
%

\begin{Lemma}\label{ste} All eigenvalues of $(S_2+\theta_0\theta_0^\t)^{-1}(\Omega_0+\theta_0\theta_0^\t)$ fall into the interval $(0,1)$.
\end{Lemma}
\noindent{\bf Proof} By
the Cauchy-Schwarz Inequality that for any $x\in R^d$,
\beginn
&&E\{g(X)(X-x)|X^\t\vartheta=x^\t\vartheta\}E\{g(X)(X-x)|X^\t\vartheta=x^\t\vartheta\}^\t\\
&&\le
E\{g(X)|X^\t\vartheta=x^\t\vartheta\}E\{g(X)(X-x)(X-x)^\t|X^\t\vartheta=x^\t\vartheta\},
\endn
which is equivalent to
\beginn
\{\nu_{\vartheta}(x)-x\mu_{\vartheta}(x)\}\{\nu_{\vartheta}(x)-x\mu_{\vartheta}(x)\}^\t\le \mu_{\vartheta}(x)\omega_{\vartheta}({x})\\
\mbox{or }\ \mu_{\vartheta}(x)\{(\nu/\mu)_{\vartheta}(x)-x\}\{(\nu/\mu)_{\vartheta}(x)-x\}^\t\le \omega_{\vartheta}({x}).
\endn
Multiply both sides by $m'(x^\theta_0)^2$ and take expectation, we have that $S_2- \Omega_0\ge 0$, which could be strengthen as
$S_2- \Omega_0> 0$. This is because if there
exists some $\vartheta_1\ne 0$, such that $\vartheta_1^\t (S_2-\Omega_0)
\vartheta_1= 0$, then for any $x,$ there exists some $C$, such that
 \beginy
\n&&\{g(X)\}^{1/2}\vartheta_1^\t (X-x)\equiv C \{g(X)\}^{1/2}, \mbox{
for all } X^\t\vartheta=x^\t\vartheta \Rightarrow \\
&&\vartheta_1^\t (X-x)\equiv C , \mbox{
for all } X^\t\vartheta=x^\t\vartheta\Rightarrow\vartheta_1\equiv
 \vartheta\label{im}
\endy

A sufficient condition for $(S_2+\theta_0\theta_0^\t)^{-1}(\Omega_0+\theta_0\theta_0^\t)$
 to have only  positive eigenvalues is  that $\theta_0$ is
the sole  eigenvector of $S_2$ and $\Omega_0$ that corresponds to eigenvalue $0$.
We argue this by contradiction. Suppose there exists some $\vartheta$ such that $\vartheta\bot\theta_0$
and
 \beginy
 &&E\{g(X)\vartheta^\t (X-x) (X-x)^\t \vartheta|\theta_0^\t X=\theta_0^\t x\}=0,
 \mbox{ for any }x\in R^d \label{rrr}\\
&&E\{g(X)\vartheta^\t (X-x)|\theta_0^\t X=\theta_0^\t x\}=0,
 \mbox{ for any }x\in R^d \label{rrr1}
 \endy
 Note that as $g(X)>0$,  (\ref{rrr}) in fact implies that $E\{\vartheta^\t (X-x)|\theta_0^\t X=\theta_0^\t x\}=0$,
 which in turn means that $\vartheta=\theta_0$; this contradicts the fact that $\vartheta\bot\theta_0$.

  To show that (\ref{rrr1}) can't be true, let $\{b_1,\cdots,b_{d-1}\}$  constitute the  orthogonal  basis
    of the orthogonal space  to vector $\theta_0$. Let $x=b_i,\ i=1,\cdots,d-1$, then $\theta_0^\t x=0$
    and from (\ref{rrr1}) we have
   \beginn
   E\{g(X)\vartheta^\t (X-b_i)|\theta_0^\t X=0\}=0,\Rightarrow
   \vartheta^\t E\{g(X)X|\theta_0^\t X=0\}= \vartheta^\t b_iE\{g(X)|\theta_0^\t X=0\}
   \endn
As $ E\{g(X)X|\theta_0^\t X=0\}$ and $E\{g(X)|\theta_0^\t X=0\}$ are constants (vector) independent of $b_i$ and
$ E\{g(X)X|\theta_0^\t X=0\}\bot \theta_0$, we have that there exists some vector $b\bot \theta_0$ such that
\beginn
\vartheta^\t b=\vartheta^\t b_i, \quad i=1,\cdots,d-1,\Leftrightarrow \vartheta^\t (b- b_i)=0 \quad i=1,\cdots,d-1,
\endn
but this can not be true unless $\vartheta\bot b_i$ for all $i=1,\cdots,d-1.$

 Next we show that $\lambda_{max}< 1$ by contradiction.
 If not, suppose $x$ is the corresponding eigenvector,
 \beginn
&&(S_2+\theta_0\theta_0^\t)^{-1}(\Omega_0+\theta_0\theta_0^\t)
x=\lambda_{max} x \Rightarrow (\Omega_0+\theta_0\theta_0^\t)
x=\lambda_{max}(S_2+\theta_0\theta_0^\t) x\\
&& \Rightarrow x^\t(\Omega_0+\theta_0\theta_0^\t)
x=\lambda_{max}x^\t (S_2+\theta_0\theta_0^\t) x \Rightarrow
x^\t\Omega_0 x\ge \lambda_{max}x^\t S_2 x(\because\lambda_{max}x\ge
1)
 \endn
which contradicts the fact that $S_2-\Omega_0>0$ if $x\ne \theta_0.$

\

\begin{center}
 {{\large R}EFERENCES}
\end{center}

\begin{description}

\ditem Andrews, D.W.K. (1994)  Asymptotics for semiparametric
econometric models via stochastic equicontinuity. {\it Econometrica}
{\bf 62}, 43-72.

\ditem Bosq, D. (1998) {\it Nonparametric Statistics for Stochastic
Processes: Estimation and Prediction}. Springer, New York.

\ditem Chaudhuri, P., Doksum, K. and Samarov, A. (1997) On average
derivative quantile regression. \textit{Ann. Statist.} \textbf{ 25},
715-44.

%\ditem Fan, Y. and Li, Q. (1999) Central Limit Theorems for
%Degenerate U-statistics of Absolutely Regular Processes With
%Applications to Model Specification Testing. \textit{Journal of
%Nonparametric Statistics} {\bf 10} 245-71.

%\ditem Gin\'{e}, E. ,Kwapi{e}\'{n}, S., Latala,  R. and Zinn, J.
%(2001)
% The LIL for canonical U-statistics of order 2. {\it
% Ann. Probab.} {\bf 29}, 520-57.

 %\ditem Ibragimov, I.A. and Linnik, Y.V. (1971) Independent and stationary
%  sequences of random variables. Wolters-Noordhoff Publishing: Groningen

\ditem Koenker, R. (2005) \textit{Quantile Regression}. Cambridge
University Press, New York.

\ditem Koenker, R. and  Bassett, G. (1978) Regression Quantiles.
Econometrica.  {\bf 46} (1),  33–50.

\ditem Kong, E., Linton, O. and Xia, Y. (2008) Uniform Bahadur
representation for local polynomial estimates of M-regression and
its application to the additive model, \textit{Econometric Theory}
(to appear)

\ditem Korolyuk, V. S.  and Borovskikh, Yu. V. (1989) Law of the
iterated logarithm for U-statistics. {\it Ukrainian Mathematical
Journal} {\bf 1}, 89-92.

\ditem Lange, T.  Rahbek, A. and Jensen, S.T. (2006) Estimation and
Asymptotic Inference in the First Order AR-ARCH Model. Preprint

\ditem Pollard, D. (1991) Asymptotics for least absolute deviation
regression estimators. {\it Econometric Theory} {\bf 7}, 186-99.

\ditem Sun, S. and Chiang, C.Y. (1997) Limiting behavior of the
perturbed empirical distribution functions evaluated at U-statistics
for strongly mixing sequences of random variables.  Journal. Applied
Mathematics and Stochastic Analysis. {\bf 10},  3-20.

 \ditem Welsh, A.H. (1996) Robust estimation of smooth regression
and spread functions and their derivatives {\it Statistica Sinica}
{\bf 6}, 347-366.

\ditem Xia, Y. (2007) Direct estimation of the multiple-index model

\end{description}

\end{document}